\documentclass[aps,pra,groupedaddress,showpacs]{revtex4}
\usepackage{color}
\usepackage{longtable}
\usepackage{dcolumn}
\usepackage{graphicx}
\usepackage{bm}
\usepackage{bbm}
\usepackage{times}
\usepackage{nicefrac}
\usepackage{amsmath}
\usepackage{amsfonts}
\usepackage{amssymb}
\usepackage{amsthm}
\usepackage{dcolumn}
\newcommand*{\centt}[1]{\multicolumn{1}{c}{#1}}

\newcolumntype{w}[1]{D{.}{.}{#1}}
\newcommand{\Za}{Z \alpha}

\begin{document}
\preprint{Version 1.0}

\title{Three-photon exchange nuclear structure correction \\ in hydrogenic systems}

\author{Krzysztof Pachucki}
\email[]{krp@fuw.edu.pl}
\homepage[]{www.fuw.edu.pl/~krp}

\affiliation{Faculty of Physics, Warsaw University,
             Pasteura 5, 02-093 Warsaw, Poland}

\author{Vojt\v{e}ch Patk\'o\v{s}}
\affiliation{Faculty of Mathematics and Physics, Charles University,  Ke Karlovu 3, 121 16 Prague
2, Czech Republic}

\author{Vladimir A. Yerokhin}
\affiliation{Center for Advanced Studies, Peter the Great St.~Petersburg Polytechnic University,
Polytekhnicheskaya 29, 195251 St.~Petersburg, Russia}

\date{\today}
\begin{abstract}

The complete relativistic $O(\alpha^2)$ nuclear structure correction to the energy levels of
ordinary (electronic) and muonic hydrogen-like atoms is investigated. The elastic part of the
nuclear structure correction is derived analytically. The resulting formula is valid for an
arbitrary hydrogenic system and is much simpler than analogous expressions previously reported in
the literature. The analytical result is verified by high-precision numerical calculations. The
inelastic $O(\alpha^2)$ nuclear structure correction is derived for the electronic and muonic
deuterium atoms. The correction comes from a three-photon exchange between the nucleus and the
bound lepton and has not been considered in the literature so far. We demonstrate that in the
case of deuterium, the inelastic three-photon exchange contribution is of a similar size and of
the opposite sign to the corresponding elastic part and, moreover, cancels exactly the model
dependence of the elastic part.  The obtained results affect the determination of nuclear charge
radii from the Lamb shift in ordinary and muonic atoms.
\end{abstract}


\maketitle
\section{Introduction}

The determination of the nuclear charge radii from atomic spectra is a very interesting test of the
Standard Model of fundamental interactions. The lepton universality, namely the identical
interaction strength of all leptons, ensures that the nuclear charge radii derived from the
ordinary (electronic) and the muonic atoms should be exactly the same. However, a series of
experiments on $\mu$H \cite{pohl:10} and $\mu$D \cite{pohl:16} and (still unpublished) measurements
on $\mu^3$He and $\mu^4$He \cite{pohl:priv} revealed significant discrepancies for the determined
nuclear charge radii, as compared to those derived from the corresponding electronic atoms. In
order to verify these discrepancies one should carefully examine all possible sources of
uncertainties in the spectroscopic determinations of the nuclear charge radii.

The main theoretical uncertainty of the Lamb shift in light muonic atoms comes from our
insufficient knowledge of the nuclear internal structure. The nuclear structure corrections are
usually divided into the elastic and the inelastic part. The elastic part (also referred to as the
finite nuclear size correction) is induced by a static distribution of the nuclear charge  and can
be obtained by solving the Dirac equation. The inelastic nuclear correction  is much more
complicated; it encompass the nuclear dipole polarizability and higher-order contributions. To deal
with the nuclear corrections, one performs an expansion of the binding energy in powers of the fine
structure constant $\alpha$ and examines the expansion terms one after another.

The leading nuclear effect is of order $\alpha^4$ and of pure elastic origin. The first-order
$O(\alpha)$ nuclear-structure correction (often referred to as the two-photon exchange
contribution) has both elastic and inelastic parts and was extensively studied both for the
electronic and the muonic atoms \cite{friar:97:b, pachucki:11:mud, pachucki:15:mud, hernandez:14,
hernandez:18}. One of the interesting results was a significant cancellation between the elastic
and the inelastic $O(\alpha)$ nuclear contributions.

The next-order $O(\alpha^2)$ nuclear structure correction comes from the three-photon exchange
between the bound lepton and the nucleus. Only the elastic part of this correction has been
addressed in the literature so far \cite{friar:79:ap}. In the present work we demonstrate that the
inelastic $O(\alpha^2)$ contribution is significant and partially cancels its elastic counterpart.
We also derive formulas for the complete $O(\alpha^2)$ nuclear correction in deuterium. Our
calculation is performed in the nonrecoil limit and neglects the magnetic dipole and electric
quadrupole moments of the nucleus. The results obtained affect determinations of nuclear charge
radii from the precision spectroscopy of ordinary and muonic atoms. However, they are not able to
explain the previously reported discrepancy between the H-D and $\mu$H-$\mu$D isotope shift
\cite{pohl:16}.

We now introduce notations for the nuclear radii that will be extensively used throughout this
paper. $r_C$ denotes the root mean square (rms) charge radius of an arbitrary nucleus, $r_C \equiv
\sqrt{\langle r^2\rangle}$. We will use specific notations for several important nuclei: $r_C({\rm
H}) \equiv r_p$ for the rms radius of the proton, $r_C({\rm D}) \equiv r_d$ for the rms radius of
the deuteron, and $r_s$ for the deuteron structure radius $r^2_s = r^2_d - r^2_p$. Since we neglect
the finite nuclear mass effects, there is no  $3/(4\,m_p^2)$ term in $r_d^2$. We define $r_{CC} =
\sqrt[\leftroot{-2}\uproot{2} 4]{\langle r^4 \rangle}$ for an arbitrary nucleus, with the specific
cases of $r_{CC}({\rm H}) \equiv r_{pp}$ for the proton, $r_{CC}({\rm D}) \equiv r_{dd}$ for the
deuteron, and $r_{ss}$ for the corresponding structure radius of the deuteron. $r_Z$ is the third
Zemach moment defined below by Eq.~(\ref{rZ}). We will also introduce two new effective nuclear
radii of arbitrary nuclei, $r_{C1}$ and $r_{C2}$, defined by Eqs.~(\ref{rC1}) and (\ref{rC2}),
respectively. The corresponding specific notations are $r_{C1}({\rm H}) \equiv r_{p1}$ and
$r_{C2}({\rm H}) \equiv r_{p2}$ for the proton and $r_{d1}$ and $r_{d2}$ for the deuteron,
respectively.

\section{Leading finite nuclear size correction}
In this section we rederive well-known results for the leading nuclear correction of order
$\alpha^4$, which is of pure elastic (finite nuclear size) origin and induced by the one-photon
exchange between the bound lepton and the nucleus. This derivation sets the ground for our further
evaluation of higher-order corrections.

Let us assume that the nucleus is a scalar particle with the charge density $\rho(q^2)$ in the
momentum space. The electron-nucleus interaction potential in momentum space is then
\begin{equation}
  V(q^2) = -\rho(q^2)\,\frac{4\,\pi\,Z\,\alpha}{q^2}\,. \label{01}
\end{equation}
The expansion coefficients of $\rho$ in $q^2$,
\begin{equation}
  \rho(q^2) = 1+\rho'(0)\,q^2 + \frac{1}{2}\,\rho''(0)\, q^4 + \ldots\,, \label{02}
\end{equation}
can be interpreted in terms of momenta of the nuclear charge distribution $\langle r^2\rangle$ and
$\langle r^4\rangle$,
\begin{eqnarray}
\rho'(0) &=& -\frac{\langle r^2\rangle}{6}, \label{03}\\
\rho''(0) &=& \frac{\langle r^4\rangle}{60}\,. \label{04}
\end{eqnarray}
From the second term in the right-hand side of Eq.~(\ref{02}), one immediately obtains the leading
finite nuclear size correction to the potential,
\begin{equation}
  \delta V = -\rho'(0)\,4\,\pi\,Z\,\alpha\,\delta^{(3)}(r)\,, \label{05}
\end{equation}
and to the energy level of a hydrogenic system,
\begin{equation}
  E^{(4)}_{\rm fns} = \langle\phi|\delta V|\phi\rangle =
  \frac{2\,\pi}{3}\,Z\,\alpha\,\langle r^2\rangle\,\phi^2(0)\,, \label{06}
\end{equation}
where, for $nS$ states,
\begin{equation}
\phi^2_{nS}(0) = \bigl< \delta^{(3)}(r)\bigr> = \frac{(\mu Z\alpha)^3}{\pi n^3}\,,
\end{equation}
and $\mu = m/(1+m/M)$ is the reduced mass of the atom.

In order to establish the importance of higher-order effects, we will need numerical values of the
leading finite nuclear size effect in hydrogen and deuterium. The corresponding results, obtained
assuming $r_p = 0.840\,87$ fm and $r_d = 2.125\,62$~fm, are, for the electronic atoms,
\begin{eqnarray}
  E^{(4)}_{\rm fns}(2S-1S,H)     &=& -1\,368\,396\,r_p^2 = \;\;-9\,67\,541\, {\rm Hz}\,, \label{07}\\
  E^{(4)}_{\rm fns}(2S-1S,D)     &=& -1\,369\,513\,r_d^2 = -6\,187\,818\,{\rm Hz}\,, \label{08}
\end{eqnarray}
and for the muonic atoms,
\begin{eqnarray}
  E^{(4)}_{\rm fns}(2P-2S,\mu H) &=& -5.197\,45\, r_p^2 = \;\;-3.674\,92\,{\rm meV}\,, \label{09}\\
  E^{(4)}_{\rm fns}(2P-2S,\mu D) &=& -6.073\,18\, r_d^2 = -27.440\,22\,{\rm meV}\,. \label{10}
\end{eqnarray}

We also observe that one of the relativistic $O(\alpha^2)$ corrections comes from the third term in
Eq.~(\ref{02}),
\begin{equation}
\delta^{(2)} V = \frac{1}{2}\,\rho''(0)\,4\,\pi\,Z\,\alpha\,\nabla^2\delta^{(3)}(r)\,. \label{11}
\end{equation}
Since its expectation value on $nS$ states is singular, we will use dimensional regularization and
combine this part with other $O(\alpha^2)$ corrections to obtain a finite result.

\section{Two-photon exchange nuclear structure: muonic atoms}
\label{sec:tpe} In this section we address the leading $O(\alpha)$ nuclear structure contribution
$E^{(5)}$ in muonic atoms, which originates from the two-photon exchange between the bound lepton
and the nucleus.

The elastic part $E^{(5)}_{\rm fns}$ can be obtained from the forward two-photon scattering
amplitude at zero momentum
\begin{equation}
  E^{(5)}_{\rm fns} = \phi^2(0)\,\int\frac{d^3q}{(2\,\pi)^3}\,
  {\rm Tr}\biggl[\gamma^0\,V\,\frac{1}{\not\!p-m}\,\gamma^0\,V\,\frac{(I+\gamma^0)}{4}\biggr]\,, \label{12}
\end{equation}
with $V = -4\,\pi\,Z\,\alpha/q^2$ and $p=(m, \vec q)$. This leads to the so-called Friar correction
\cite{friar:79:ap},
\begin{eqnarray}
  E^{(5)}_{\rm fns} &=&
-(4\,\pi\,Z\,\alpha)^2\,\phi^2(0)\,2\,m\int\frac{d^3q}{(2\,\pi)^3}\,\frac{\rho^2(q^2)-1-2\,q^2\,\rho'(0)}{q^6}
\nonumber \\ &=&  -\frac{\pi}{3}\,\phi^2(0)\,(Z\,\alpha)^2\,m\,r_Z^3\,,\label{13}
\end{eqnarray}
where
\begin{equation} \label{rZ}
r_Z^3 =  \int d^3r_1\int d^3r_2\,\rho(r_1)\,\rho(r_2)\,|\vec r_1-\vec r_2|^3\,.
\end{equation}
As pointed out in Refs.~\cite{friar:97:b, pachucki:11:mud}, it is important to consider the Friar
correction $E^{(5)}_{\rm fns}$ together with the corresponding inelastic part, because of a
cancellation between them, occurring both for the muonic and the ordinary atoms. For this reason,
we do not separate out $E^{(5)}_{\rm fns}$ but absorb it in the total nuclear structure correction
$E^{(5)}$.

\subsection{Muonic hydrogen}

The inelastic two-photon exchange correction in $\mu$H has been extensively studied in the
literature (see Ref.~\cite{birse:2012} and references therein). It is also given by the forward
scattering amplitude and can be parameterized in terms of two spin-independent structure functions
of the proton. Using dispersion relations, these functions are usually expressed in terms of the
cross section of the inelastic photon scattering off the proton, which is extracted from
experiment. The main problem of this approach is that one of the dispersion relations involves
subtractions that can only be obtained from theory, and this introduces the dominant uncertainty.

There is good agreement between different calculations of the two-photon exchange correction, with
the final result of $E^{(5)} (2P_{1/2}\mbox{\rm -- }2S, \mu{\rm H}) = E^{(5)}_{\rm fns} +
E^{(5)}_{\rm pol} = 0.033\,2(20)$~meV assumed by the CREMA collaboration \cite{antognini:13:ap} in
their determination of the proton charge radius. It is convenient to parameterize this result in
terms of an effective radius $r_{pF}$, in analogy to Eq.~(\ref{13}),
\begin{equation}
  E^{(5)}(\mu {\rm H}) =  -\frac{\pi}{3}\,\phi^2(0)\,(Z\,\alpha)^2\,m\,r_{pF}^3\,, \label{15}
\end{equation}
with
\begin{equation}
  r_{pF}^3 = 3.270\,(197) \;{\rm fm}^3\,. \label{16}
\end{equation}
This parametrization will be used below in our calculation of the inelastic contribution in other
muonic atoms, see Eq.~(\ref{23}).

\subsection{Muonic atoms other than hydrogen}

For all nuclei other than the proton, the inelastic contribution is dominated by the electric
dipole polarizability. For muonic atoms, one may assume the nonrelativistic approximation, so the
second-order correction due to the electric dipole nuclear excitation is
\begin{equation}
E^{(5)}_{\rm pol0} = \alpha^2\,\biggl\langle\!\phi\,\phi_N\biggl|
\frac{\vec d\cdot\vec r}{r^3}\,
\frac{1}{E_N + E_0 - H_N -H_0}\,
\frac{\vec d\cdot\vec r}{r^3}\,
\biggr|\phi\,\phi_N\!\biggr\rangle\,, \label{17}
\end{equation}
where $\vec d$ is the electric dipole operator divided by the elementary charge, and $H_0$ and
$H_N$ are the nonrelativistic Coulomb Hamiltonian for the muon and the nucleus, respectively. To
the leading order in $\alpha$, one may neglect the Coulomb interaction and replace
$\phi(r)\rightarrow \phi(0)$ to obtain a compact formula for the leading two-photon exchange
contribution,
\begin{equation}
  E^{(5)}_{\rm pol0} = -\frac{4\,\pi\,\alpha^2}{3}\,\phi^2(0)
  \biggl\langle\phi_N\biggl|\vec d\,\sqrt{\frac{2\,m}{H_N-E_N}}\vec d\,\biggr|\phi_N\biggr\rangle\,,
  \label{18}
\end{equation}
which contributes $1.910$ meV to the $2P-2S$ transition energy in muonic deuterium
\cite{pachucki:15:mud}.

There are many corrections to the leading contribution \cite{pachucki:11:mud, pachucki:15:mud,
hernandez:14, hernandez:18}, the most interesting of them being the one that partially cancels the
Friar correction. To show this, following Ref. \cite{pachucki:11:mud}, we consider the muonic
matrix element $P$ for the nonrelativistic two-photon exchange
\begin{equation}
P =\sum_{i,j}\biggl\langle\phi\biggl|\frac{\alpha}{|\vec r-\vec R_i|}
\frac{1}{(H_0-E_0+E)} \frac{\alpha}{|\vec r-\vec R_j'|}\biggr|\phi\biggr\rangle, \label{19}
\end{equation}
where $H_0$ is the nonrelativistic Hamiltonian for the muon (electron) in the nonrecoil limit, and
$\vec R_i$ is a position of the $i$th proton with respect to the nuclear mass center. Using the
on-mass-shell approximation, subtracting  the leading Coulomb interaction, the finite nuclear size,
and the electric dipole polarizability, and expanding in the small parameter $\sqrt{2\,m\,E} |\vec
R_i-\vec R_j'|$, we obtain
\begin{eqnarray}
P &=& \alpha^2\,\phi^2(0)\,
\int\frac{d^3 q}{(2\,\pi)^3}\,\biggl(\frac{4\,\pi}{q^2}\biggr)^2\,
\biggl(E+\frac{q^2}{2\,m}\biggr)^{-1}\,
\biggl[e^{i\,\vec q\cdot(\vec R-\vec R')}-1+\frac{q^2}{6}\,(\vec R-\vec R')^2\biggr]
\label{20}\\&\approx&
\sum_{i,j}\frac{\pi}{3}\,m\,\alpha^2\,\phi^2(0)\,|\vec R_i-\vec R_j'|^3
\biggl(1-\frac{1}{5}\,\sqrt{2\,m\,E}|\vec R_i-\vec R_j'| +\ldots\,\biggr)\,. \nonumber
\end{eqnarray}
The corresponding correction to the atomic energy is
\begin{equation}
E^{(5)}_{\rm pol1} = -\sum\hspace{-3.0ex}\int dE\!\!\int\!\!d^3 R\,d^3R'\,
\phi^*_N(\vec R)\,\phi_E(\vec R)\,\phi^*_E(\vec R')\,\phi_N(\vec R')\,P \label{21}
\end{equation}
Let us consider only the first, $E$-independent term. When $\phi_E = \phi_N$, it corresponds to the
elastic part, namely, the Friar correction given by Eq.~(\ref{13}). However, the inclusion of all
excited states leads to
\begin{equation}
E^{(5)}_{\rm pol1} = -\frac{\pi}{3}\,m\,\alpha^2\,\phi^2(0)\,\sum_{i,j=1}^Z\langle\phi_N||\vec R_i-\vec R_j|^3|\phi_N\rangle\,,
  \label{22}
\end{equation}
which is much different from Eq.~(\ref{13}), in particular it vanishes for deuterium.

There are further nuclear polarizability corrections which were extensively studied in the
literature \cite{pachucki:11:mud, pachucki:15:mud, hernandez:14, hernandez:18}. It is convenient to
write the final result for the two-photon exchange nuclear structure correction separating out the
contribution due to the two-photon exchange with individual nucleons,
\begin{equation}
  E^{(5)} = E^{(5)}_{\rm pol} - \frac{\pi}{3}\,m\,\alpha^2\,\phi^2(0)\,
  \biggl[Z\,r^3_{pF} + (A-Z)\,r^3_{nF}\biggr]\,, \label{23}
\end{equation}
where $E^{(5)}_{\rm pol} = E^{(5)}_{\rm pol0} + E^{(5)}_{\rm pol1} + \ldots$.  Such representation
of the nuclear structure correction is particularly advantageous for calculating isotope shifts,
since the individual nucleon contributions partially cancel each other in the difference, together
with the corresponding uncertainties. Calculating the two-photon exchange with individual nucleons,
we take the effective proton radius $r^3_{pF}$ from Eq.~(\ref{16}), whereas for the neutron we assume
the corresponding parameter to be four times smaller than that of the proton, $r_{nF}^3 =
r_{pF}^3/4$, with uncertainty of 100\%. This choice of $r_{nF}^3$ is in agreement with results
summarized in Ref.~\cite{krauth:16:ap} but requires further investigations.

Despite the fact that the literature results for $E^{(5)}_{\rm pol}$ in $\mu$D reported by
different groups \cite{pachucki:11:mud, pachucki:15:mud, hernandez:14, hernandez:18} are in good
agreement with each other (see a summary in Ref.~\cite{krauth:16:ap}), one should bear in mind that
a number of higher-order effects exist that have not yet been addressed in any of the previous
studies. Specifically, it has not so far been possible to include nucleon relativistic corrections
to the coupling of the nucleus to the electromagnetic field. We thus believe that all theoretical
predictions of $E^{(5)}_{\rm pol}$ in $\mu$D should bear an uncertainty whose relative value is 
approximately the ratio of the average nucleon binding energy to the nucleon mass, which is about 1\%.

Summarizing our analysis of the existing literature results, we adopt the sum of entries $p_1\ldots
p_{12}$ labelled as ``Our choice'' in Table~3 in Ref.~\cite{krauth:16:ap} as currently the best value of the
two-photon nuclear polarizability correction to the $2P_{1/2} \mbox{\rm --} 2S$ transition energy
in $\mu$D, and ascribe the uncertainty of 1\% to it,
\begin{equation}
  E^{(5)}_{\rm pol}(2P_{1/2} \mbox{\rm --} 2S,\mu D) = 1.6625\,(166)\, {\rm meV} \label{25}\,.
\end{equation}
The above uncertainty of $E^{(5)}_{\rm pol}$ is about 50\% larger than the corresponding
estimate of $\pm 0.0107$~meV given in Table~3 of Ref.~\cite{krauth:16:ap}. Finally, we add the
individual nucleon part in Eq.~(\ref{23}) and obtain the total two-photon nuclear structure
correction to the $2P_{1/2} \mbox{\rm --} 2S$ transition energy in $\mu$D,
\begin{equation}
  E^{(5)}(2P_{1/2} \mbox{\rm --} 2S,\mu D) = 1.7110\,(194)~{\rm meV}\,,\label{24a}
\end{equation}
which almost coincides with the corresponding result  of $1.7096\,(200)$~meV from
Ref.~\cite{krauth:16:ap}, as given by Eq.~(17) of that work.

\section{Two-photon exchange nuclear structure: electronic atoms}

The elastic (finite nuclear size) part of the two-photon exchange nuclear structure correction
for electronic atoms is given by the same formula as for the muonic atoms, Eq.~(\ref{12}).

\subsection{Electronic hydrogen}

We calculate the elastic part of the nuclear structure correction for hydrogen according to
Eq.~(\ref{12}) and using the result for the third Zemach moment from Ref.~\cite{borie:12} obtained
by averaging values measured in scattering experiments,
\begin{equation}
r_{pZ} = 1.587\,(26)\,r_p\,.
\end{equation}
The corresponding result for the $2S$--$1S$ transition is
\begin{equation}
 E^{(5)}_{\rm fns}(2S \mbox{\rm --} 1S,e{\rm H}) = 0.0307\,(15)~\mbox{\rm kHz}\,.
\end{equation}

The inelastic part of the two-photon exchange nuclear structure correction $E^{(5)}_{\rm pol}$ was
derived in the logarithmic approximation in Ref.~\cite{khriplovitch:00},
\begin{equation}
  E^{(5)}_{\rm pol}(2S \mbox{\rm --} 1S,e{\rm H}) = -m\,\alpha\,\phi^2(0)\,\bigl[5\,\alpha_p -\beta_p\bigr]\,\ln\frac{\bar E_p}{m}\,,
\end{equation}
where $\bar E_p$ is the average proton excitation energy and $\alpha_p$ and $\beta_p$ are the
static proton polarizabilities extracted from experiment. Using the same average proton excitation
energy $\bar E_p = 410$~MeV as in Ref.~\cite{khriplovitch:00} and the updated results for the
proton polarizabilities \cite{demissie:16},
\begin{eqnarray}
  \alpha_p &=& 10.65\,(35)(20)(30)\,\mbox{\rm fm}^3\,,\nonumber\\
  \beta_p &=& 3.15\,(35)(20)(30)\,\mbox{\rm fm}^3\,,
\end{eqnarray}
we obtain the result for the $2S$--$1S$ transition of
\begin{equation}
 E^{(5)}_{\rm pol}(2S \mbox{\rm --} 1S,e{\rm H}) = 0.0567\,(85)~\mbox{\rm kHz}\,,
\end{equation}
where, following Ref.~\cite{khriplovitch:00}, we assumed a 15\% uncertainty due to the leading
logarithmic approximation.

The total result for the two-photon nuclear structure correction in electronic hydrogen is
\begin{equation}
E^{(5)}(2S \mbox{\rm --} 1S,e{\rm H})  = E^{(5)}_{\rm fns} + E^{(5)}_{\rm pol} = 0.0874\,(86)~\mbox{\rm kHz}\,,
\label{E5H}
\end{equation}
which could be compared with the corresponding result of $0.091\,(11)$~kHz from
Ref.~\cite{mohr:16:codata}.

\subsection{Electronic atoms other than hydrogen}

Similarly to the muonic atoms, it is convenient to write the total two-photon exchange nuclear
structure correction separating out the contribution due to the interaction with individual
nucleons,
\begin{equation}
  E^{(5)} = E^{(5)}_{\rm pol} - m\,\alpha\,\phi^2(0)\,\biggl[\frac{\pi}{3}\,\alpha\,Z\,r^3_{pZ}
  +Z\,\bigl[5\,\alpha_p -\beta_p\bigr]\,\ln\frac{\bar E_p}{m}
   +(A-Z)\,\bigl[5\,\alpha_n -\beta_n\bigr]\,\ln\frac{\bar E_n}{m}\biggr]\,. \label{23b}
\end{equation}
In the above formula, the first and the second terms in the brackets represent the elastic and the
inelastic interactions with individual protons, respectively, whereas the third term comes from the
inelastic interaction with individual neutrons. The parameters for the protons are the same as for
hydrogen, whereas for the neutrons we use the experimental polarizabilities \cite{demissie:16},
\begin{eqnarray}
  \alpha_n &=& 11.55\,(125)(20)(80)\,\mbox{\rm fm}^3\,,\nonumber \\
  \beta_n &=& 3.65\,(125)(20)(80)\,\mbox{\rm fm}^3\,,
\end{eqnarray}
and the same value of $\bar E_n = 410$~MeV as for the proton. We note that the elastic interaction
of the bound electron with the nucleus as a whole is absorbed in $E^{(5)}_{\rm pol}$, reflecting
the fact that the third Zemach moment correction for a compound nucleus largely cancels out between
the elastic and inelastic parts in the same way as in muonic atoms.

Similarly to the muonic atoms, the nuclear polarizability correction $E^{(5)}_{\rm pol}$ in
Eq.~(\ref{23b}) comes from the electric dipole polarizability, which, however, takes a very
different form for the electronic atoms. Since in this case the nonrelativistic approximation is
not valid, one should consider the complete two-photon exchange and keep the relativistic form of
the matrix elements,
\begin{eqnarray}
E_{\rm pol}^{(5)} &=&
i\,e^2\,\phi^2(0)\,
\int\frac{d\,\omega}{2\,\pi}\,\int\frac{d^3k}{(2\,\pi)^3}\,\omega^2\,
\frac{\Bigl(\delta^{ik}-\frac{k^i\,k^k}{\omega^2}\Bigr)}{\omega^2-k^2}\,
\frac{\Bigl(\delta^{jl}-\frac{k^j\,k^l}{\omega^2}\Bigr)}{\omega^2-k^2}
\nonumber \\ && \times
{\rm Tr}\biggl[\biggl(\gamma^j\,\frac{1}{\not\!p-\not\!k-m}\gamma^i +
\gamma^i\frac{1}{\not\!p+\not\!k-m}\,\gamma^j\biggr)\,\frac{(\gamma^0+I)}{4}\biggr]\,
\nonumber \\ && \times
\biggl\langle\phi_N\biggl|d^{\,k}\,\frac{1}{E_N-H_N-\omega}\,
d^{\,l}\biggr|\phi_N\biggr\rangle + \ldots \nonumber\\
 &=& E^{(5)}_{\rm pol1} + E^{(5)}_{\rm pol2} +  E^{(5)}_{\rm pol3} +\ldots \label{epol}\,,
\end{eqnarray}
where $p=(m,\vec 0)$. Assuming that the electron mass is much smaller than the nuclear excitation
energy, the leading nuclear polarizability correction becomes
\begin{equation}
  E^{(5)}_{\rm pol1} = -m\,\alpha^2\,\phi^2(0)\,\frac{2}{3}\,
  \biggl\langle\phi_N\biggl|\vec d\,\frac{1}{H_N-E_N}\,
  \biggl[\frac{19}{6} + 5\,\ln\frac{2\,(H_N-E_N)}{m} \biggr]\,\vec d\,\biggr|\phi_N\biggr\rangle\,. \label{epol1}
\end{equation}
The corresponding contribution to the $2S$--$1S$ transition in ordinary deuterium is
$19.26\,(6)$~kHz \cite{friar:97:a}.

Various small corrections to the electric dipole polarizability for electronic atoms were
considered by Friar in \cite{friar:97:b}. In particular, it was shown there that the Zemach
contribution for deuterium vanishes in the same way as for the muonic deuterium. Furthermore, the
higher-order terms in the $m/(H_N-E_N)$ expansion of Eq.~(\ref{epol}) give rise to a correction
\begin{equation}
  E^{(5)}_{\rm pol2} =
    -m^3\,\alpha^2\,\phi^2(0)\,\frac{2}{3}\,
  \biggl\langle\phi_N\biggl|\vec d\,\frac{1}{(H_N-E_N)^3}\,
  \biggl[-\frac{283}{80} + \frac{15}{4}\,\ln\frac{2\,(H_N-E_N)}{m} \biggr]\,\vec d\,\biggr|\phi_N\biggr\rangle\,,
\label{epol2}
\end{equation}
which contributes 0.106~kHz to the $2S$--$1S$ transition in ordinary deuterium \cite{friar:97:b}.
Another important correction is the one due to the magnetic suceptibility \cite{friar:97:b},
\begin{equation}
  E^{(5)}_{\rm pol3} =
    m\,\alpha^2\,\phi^2(0)\,\frac{2}{3}\,
  \biggl\langle\phi_N\biggl|\vec \mu\,\frac{1}{(H_N-E_N)'}\,
  \biggl[-\frac{1}{6} + \ln\frac{2\,(H_N-E_N)}{m} \biggr]\,\vec \mu\,\biggr|\phi_N\biggr\rangle\,,
\label{epol3}
\end{equation}
where $\mu$ is the magnetic moment operator divided by an elementary charge.
It leads to a correction of $-0.307(2)(6)$~kHz to the
$2S$--$1S$ transition in $e$D \cite{friar:97:b}.

There were further corrections to the electric dipole polarizability considered in
Ref.~\cite{friar:97:b}. However, we are convinced that they were not treated correctly and,
moreover, that there are many more relativistic corrections of the same order. For this reason we
disregard the additional corrections from Ref.~\cite{friar:97:b} and assume the total
polarizability correction to be the sum of Eqs. (\ref{epol1}), (\ref{epol2}), and (\ref{epol3}).
Specifically, the result for the nuclear polarizability to the $2S$--$1S$ transition in electronic
deuterium is
\begin{equation}
E_{\rm pol}^{(5)}(2S\mbox{\rm --} 1S,D) = 19.06\,(20)\,{\rm kHz}\,. \label{epolnum}
\end{equation}
Adding the individual nucleon part contribution of $0.149\,(22)$~kHz, we obtain the total
two-photon exchange nuclear structure contribution of
\begin{equation}
E^{(5)}(2S\mbox{\rm --} 1S,D) = 19.21\,(20)\,{\rm kHz}\,,
\label{E5D}
\end{equation}
which could be compared with the sum of the nuclear polarizability correction and the third Zemach
contribution from Ref.~\cite{mohr:16:codata}, $18.70\,(7) + 0.51 = 19.21\,(7)$~kHz, perfect
agreement of the numerical values being probably accidental.

\section{Three-photon exchange elastic contribution}
This contribution has been studied by different methods and a number of authors, of note
analytically by Friar in Ref. \cite{friar:79:ap}, and numerically by solving the Dirac equation in
the field of finite size nucleus \cite{indelicato:13}. Here we present an alternative analytical
approach, which leads to much simpler analytic formulas. A numerical verification of our formulas
is given in Appendix \ref{app:2}.

In the standard analytic approach, one applies the perturbation theory to the Dirac energies with
the perturbing potential $\delta V = V - V_0$ where $V_0 = -Z\,\alpha/r$ and $V$ is the Coulomb
potential from the finite size nucleus,
\begin{eqnarray}
  \delta^{(1)}E &=& \langle\bar\psi|\delta V|\psi\rangle \label{26}\\
  \delta^{(2)}E &=& \langle\bar\psi|\delta V\,\frac{1}{(\not\!p-\gamma^0\,V_0-m)'}\,\delta V|\psi\rangle \label{27}\\
  \delta^{(3)}E &=& \langle\bar\psi|\delta V\,\frac{1}{(\not\!p-\gamma^0\,V_0-m)'}\,\delta V\,
  \frac{1}{(\not\!p-\gamma^0\,V_0-m)'}\,\delta V|\psi\rangle \label{28}
\end{eqnarray}
One can use the exact Dirac wave function and the reduced Dirac propagators to calculate the
$O(\alpha^2)$ correction to the finite nuclear size \cite{friar:79:ap}, which we call the elastic
three-photon exchange correction. However, we will not use the above formulas but employ a
different approach, which we call the scattering amplitude approach. In this approach, the
$O(\alpha^2)$ relativistic correction to the finite nuclear size is induced by the elastic three
photon exchange. The corresponding correction can be divided into the low and the high energy
momentum exchange parts, $E^{(6)}_{\rm fns} = E_L + E_H$. These parts are calculated as follows.

\subsection{Three-photon exchange: low energy part}
The low-energy part $E_L$ is again split into two parts
\begin{eqnarray}
  E_L &=& E_{L1} + E_{L2} \label{29}\,,\\
   E_{L1} &=& \langle\delta V\,\frac{1}{(E-H)'}\,\delta V\rangle  + \langle\delta^{(2)} V\rangle\,, \label{30}\\
   E_{L2} &=& \langle\phi|\frac{1}{8\,m^2}\,\nabla^2(\delta V) +
   \frac{1}{4\,m^2}\,\vec\sigma\cdot\vec \nabla(\delta V)\times \vec p|\phi\rangle
\nonumber \\ &&
  + 2\, \langle\phi|\delta V\frac{1}{(E-H)'}\,\biggl[-\frac{p^4}{8\,m^3} +\frac{\pi\,Z\,\alpha}{2\,m^2}\,\delta^3(r)
   \biggr]|\phi\rangle\,, \label{31}
\end{eqnarray}
where $E_{L1}$ is the nonrelativistic contribution proportional to $r_C^4$, and $E_{L2}$ is the
relativistic part proportional to $r_C^2$. All these matrix elements are calculated in
$d=3-2\,\epsilon$ dimensions. The following results are obtained for the $nS$ states,
\begin{eqnarray}
  E_{L1}(nS) &=& [\rho'(0)]^2\,\frac{16\,(Z\,\alpha)^6}{n^3}\biggl[-\frac{1}{n} -\frac12 + \gamma -\ln\frac{n}{2}+\Psi(n)\biggr]\nonumber \\
            && +(Z\,\alpha)^3\,
  \biggl[-\frac{1}{\epsilon} + 4\,\ln(Z\,\alpha)\biggr]\,4\,[\rho'(0)]^2\,\langle \pi\,\delta^{(d)}(r)\rangle+
  \frac{4\,(Z\,\alpha)^6\,\rho''(0)}{n^5} \,,\label{32}\\
  E_{L2}(nS) &=& \rho'(0)\,\frac{4\,(Z\,\alpha)^6}{n^3}\biggl[\frac{9}{4n^2}-\frac{1}{n}-\frac52 +\gamma -\ln\frac{n}{2}+\Psi(n)\biggr]\nonumber \\
            && +(Z\,\alpha)^3\,
  \biggl[-\frac{1}{\epsilon} + 4\,\ln(Z\,\alpha)\biggr]\,\rho'(0)\,\langle \pi\,\delta^{(d)}(r)\rangle \,, \label{33}
\end{eqnarray}
and for the $nP$ states,
\begin{eqnarray}
  E_L(nP_{1/2}) &=&(Z\,\alpha)^6\,\biggl( -\frac{9}{4}\,\rho'(0) + 3\,\rho''(0)\biggr)\,R'_{n1}(0)^2 \,,\label{34}\\
  E_L(nP_{3/2}) &=&(Z\,\alpha)^6\,3\,\rho''(0)\,R'_{n1}(0)^2 \,,\label{35}
\end{eqnarray}
where
\begin{equation}
  R'_{n1}(0)^2 = \frac{4}{9\,n^3}\,\biggl(1-\frac{1}{n^2}\biggr)\,. \label{36}
\end{equation}
For all higher-$L$ states $E_L$ vanishes.

\subsection{Three-photon exchange: high energy part}
We start by introducing the two potentials in $d$-dimensions that will appear in the evaluation of
the high energy-part $E_H$,
\begin{eqnarray}
  V_d(r) &=& 4\,\pi\,\int\frac{d^dq}{(2\,\pi)^d}\,e^{i\vec{q}\cdot\vec{r}}\,\frac{\rho(q^2)}{q^2}\,, \label{37}\\
  V_d^{(2)}(r) &=& 4\,\pi\,\int\frac{d^dq}{(2\,\pi)^d}\,e^{i\vec{q}\cdot\vec{r}}\,\frac{\rho(q^2)}{q^4}\,. \label{38}
\end{eqnarray}
Their large $r$ asymptotics are
\begin{eqnarray}
  V_d(r) & = & {\cal V}(r) + \mbox{\rm local terms}\,, \label{39}\\
  V_d^{(2)}(r) & = &  {\cal V}^{(2)}(r) + \rho'(0)\,{\cal V}(r) + \mbox{\rm local terms}\,, \label{40}
\end{eqnarray}
and in $d=3$,
\begin{eqnarray}
  V(r) & = & \frac{1}{r} + \mbox{\rm local terms}\,, \label{41}\\
  V^{(2)}(r) & = & -\frac{r}{2} + \frac{\rho'(0)}{r} + \mbox{\rm local terms}\,, \label{42}
\end{eqnarray}
where the local terms vanish outside the nucleus.

Now we proceed to the derivation of the high-energy part $E_H$. It is given by the three-photon
scattering amplitude with momenta $p_i = (m,\vec q_i)$,
\begin{eqnarray}
  E_H &=& -(4\,\pi\,Z\,\alpha)^3\,\phi^2(0)\,\int \frac{d^dq_1}{(2\,\pi)^d}\,\int \frac{d^dq_2}{(2\,\pi)^d}\,
  \frac{\rho(q_1^2)}{q_1^4}\, \frac{\rho(q_2^2)}{q_2^4}\, \frac{\rho((\vec q_1-\vec q_2)^2)}{(\vec q_1-\vec q_2)^2}\nonumber \\ &&\times
       {\rm Tr}\biggl[(\not\!p_1+m)\,\gamma_0\, (\not\!p_2+m)\, \frac{(\gamma_0+I)}{4}\biggr]\,. \label{43}
\end{eqnarray}
The above trace equates to $ 4\,m^2 +\vec q_1\,\vec q_2$, so we can split $E_H$ into the
nonrelativistic and relativistic parts,
\begin{equation}
  E_H  = E_{H1} + E_{H2} \label{44}\,.
\end{equation}
The nonrelativistic part $E_{H1}$ is
\begin{eqnarray}
  E_{H1} &=& -(4\,\pi\,Z\,\alpha)^3\,\phi^2(0)\,4\,m^2\!\int\!\frac{d^dq_1}{(2\,\pi)^d}\int\!\frac{d^dq_2}{(2\,\pi)^d}\,
 \frac{\rho(q_1^2)}{q_1^4}\, \frac{\rho(q_2^2)}{q_2^4}\, \frac{\rho((\vec q_1-\vec q_2)^2)}{(\vec q_1-\vec q_2)^2} \nonumber \\&=&
 -\phi^2(0)\,(Z\,\alpha)^3\,4\,m^2\int d^dr \, V_d(r)\,\left[V_d^{(2)}(r)\right]^2 \label{45}\,.
\end{eqnarray}
In order to calculate this integral, we split the integration region into $r<\Lambda$ and $r\geq
\Lambda$. The first integral is finite in $d=3$ but diverges at large $\Lambda$, and in the second
integral one can use the asymptotic form of potentials,
\begin{eqnarray}
 E_{H1} &=&
 -\phi^2(0)\,(Z\,\alpha)^3\,4\,m^2\biggl\{4\,\pi\,\int^\Lambda dr r^2\,V(r)\,\bigl[V^{(2)}(r)\bigr]^2
  +\int_\Lambda d^d r\,{\cal V}(r)\,\bigl[{\cal V}^{(2)}(r)  + \rho'(0)\,{\cal V}(r)\bigr]^2\biggr\} \nonumber \\ &=&
  \phi^2(0)\,(Z\,\alpha)^3\,4\,m^2\,4\,\pi\,\int^\infty dr \ln(r/r_{C})\,
  \frac{d}{dr}\,r^3\biggl\{V(r)\,\left[V^{(2)}(r)\right]^2 -\frac{1}{r}\,\biggl(\frac{r}{2}-\frac{\rho'(0)}{r}\biggr)^2
    \biggr\} \nonumber \\&&
  -\phi^2(0)\,(Z\,\alpha)^3\,4\,m^2\,\rho'(0)^2\,\biggl\{4\,\pi\,\ln(\Lambda/r_{C})
  +\int_\Lambda d^d r\,{\cal V}(r)^3\biggr\} \label{46}\,.
\end{eqnarray}
The expression under the first integral is a local function of $r$, so this integral is effectively
over the nuclear size, which allows us to introduce an effective nuclear radius $r_{C1}$ as
\begin{equation}\label{rC1}
\ln\frac{r_{C1}}{r_C} +2= \frac{36}{r_C^4}\,\int^\infty dr \ln(r/r_{C})\,
  \frac{d}{dr}\,r^3\biggl\{V(r)\,\left[V^{(2)}(r)\right]^2 -\frac{1}{r}\,\biggl(\frac{r}{2}-\frac{\rho'(0)}{r}\biggr)^2\biggr\}\,.
\end{equation}
So, the first $O(\alpha^2)$ correction $E_{H1}$ is represented in the following form
\begin{equation}
  E_{H1} = 4\,\pi\,\phi^2(0)\,(Z\,\alpha)^3\,4\,m^2\,\frac{r_C^4}{36}\,
    \biggl[\frac{1}{4\,\epsilon} + \frac{5}{2}+\gamma +\ln(r_{C1}\,m)\biggr]\,. \label{48}
\end{equation}
The relativistic part $E_{H2}$ is
  \begin{eqnarray}
E_{H2} &=& -(4\,\pi\,Z\,\alpha)^3\,\phi^2(0)\!\int\!\frac{d^dq_1}{(2\,\pi)^d}\int\!\frac{d^dq_2}{(2\,\pi)^d}\,
 \frac{\rho(q_1^2)}{q_1^4}\, \frac{\rho(q_2^2)}{q_2^4}\, \frac{\rho((\vec q_1-\vec q_2)^2)}{(\vec q_1-\vec q_2)^2}\, \vec q_1\vec q_2
  \nonumber \\&=&
  \phi^2(0)\,(Z\,\alpha)^3\,\int d^dr \, \biggl\{2\,\pi\,\rho(r)\,V_d^{(2)}(r) - \left[V_d(r)\right]^2\biggr\}\,V_d^{(2)}(r) \label{49}\,,
  \end{eqnarray}
  and we proceed in a similar way as in the case of $E_{H1}$, namely
\begin{eqnarray}
 E_{H2} &=&
  \phi^2(0)\,(Z\,\alpha)^3\,\biggl\{4\,\pi\,\int^\Lambda dr r^2
  \biggl[2\,\pi\,\rho(r)\,V^{(2)}(r) - \left[V(r)\right]^2\biggr]\,V^{(2)}(r)\nonumber \\&&
  -\int_\Lambda d^d r \left[{\cal V}(r)\right]^2\,\bigl({\cal V}^{(2)}(r)  + \rho'(0)\,{\cal V}(r)\bigr)\biggr\} \nonumber \\ &=&
  -\phi^2(0)\,(Z\,\alpha)^3\,4\,\pi\,\int^\infty dr \ln(r/r_{C})
  \frac{d}{dr}\,r^3\biggl[2\,\pi\,\rho(r)\,\left[V^{(2)}(r)\right]^2 - \left[V(r)\right]^2\,V^{(2)}(r)
    \nonumber \\&& - \frac{1}{r^2}\,\biggl(\frac{r}{2}-\frac{\rho'(0)}{r}\biggr)\biggr]
  -\phi^2(0)\,(Z\,\alpha)^3\,\rho'(0)\,\biggl[\,4\,\pi\,\ln(\Lambda/r_{C})
  + \int_\Lambda d^d r\,\left[{\cal V}(r)\right]^3\biggr] \label{50}\,.
\end{eqnarray}
The expression under the first integral is a local function, so we can introduce the second
effective nuclear radius $r_{C2}$ as
\begin{equation}\label{rC2}
\ln\frac{r_{C2}}{r_C}-1 = \frac{6}{r_C^2}\,\int^\infty dr \ln(r/r_{C})
  \frac{d}{dr}\,r^3\biggl[2\,\pi\,\rho(r)\,\left[V^{(2)}(r)\right]^2 - \left[V(r)\right]^2\,V^{(2)}(r)
    - \frac{1}{r^2}\,\biggl(\frac{r}{2}-\frac{\rho'(0)}{r}\biggr)\biggr]\,.
\end{equation}
So, the second $O(\alpha^2)$ correction is given by
\begin{equation}
  E_{H2} = -4\,\pi\,(Z\,\alpha)^3\,\phi^2(0)\,\frac{r_C^2}{6}\,
  \biggl[\frac{1}{4\,\epsilon} -\frac{1}{2}+\gamma + \ln(r_{C2}\,m)\biggr] \label{52}\,.
\end{equation}

\subsection{Three-photon elastic exchange: total result}
The complete $O(\alpha^2)$ finite nuclear size correction for an arbitrary nucleus is given by the
sum $E^{(6)}_{\rm fns} = E_L + E_H = E_{L1} + E_{L2} + E_{H1} + E_{H2}$, with the result
\begin{eqnarray}
E^{(6)}_{\rm fns}(nS) &=&-(Z\,\alpha)^6\,m^3\,r_C^2\,\frac{2}{3\,n^3}\,\biggl[\frac{9}{4n^2} - 3 -\frac{1}{n} +2\,\gamma -\ln\frac{n}{2}+\Psi(n)
        +\ln(m\,r_{C2}\,Z\,\alpha)\biggr]
        \nonumber \\
        && + (Z\,\alpha)^6\,m^5\,r_C^4\,\frac{4}{9\,n^3}\,\biggl[-\frac{1}{n} + 2 + 2\,\gamma -\ln\frac{n}{2}+\Psi(n)
        + \ln(m\,r_{C1}\,Z\,\alpha)\biggr]
        \nonumber \\ &&
        +(Z\,\alpha)^6\,m^5\,r_{CC}^4\,\frac{1}{15\,n^5}  \label{53}\,,\\
  E^{(6)}_{\rm fns}(nP_{1/2}) &=&(Z\,\alpha)^6\,m\,
  \biggl(\frac{m^2\,r_C^2}{6} + \frac{m^4\,r_{CC}^4}{45}\biggr)\,\frac{1}{n^3}\,\biggl(1-\frac{1}{n^2}\biggr) \label{54}\,, \\
  E^{(6)}_{\rm fns}(nP_{3/2}) &=&(Z\,\alpha)^6\,m^5\,r_{CC}^4\,\frac{1}{45\,n^3}\,\biggl(1-\frac{1}{n^2}\biggr) \label{55}\,,\\
  E^{(6)}_{\rm fns}(nL_{J}) &=& 0\;\mbox{\rm for}\; L>1\,, \label{56}
\end{eqnarray}
where $r_{CC}^4 = \langle r^4\rangle$ and the effective nuclear charge radii $r_{C1}$ and $ r_{C2}$
defined by Eqs.~(\ref{rC1}) and (\ref{rC2}) encode the high-momentum contributions and are expected
to be of the order of $r_C$. Equations~(\ref{53})-(\ref{56}) are valid both for electronic and
muonic atoms. However, in the case of the electronic atoms, the terms proportional to $r_C^4$ and
$r_{CC}^4$ in these formulas are smaller than the next-order correction and thus should be
neglected.

Equations (\ref{53})-(\ref{56}) depend on the nuclear model through the effective nuclear charge
radii $r_{C1}$ and $ r_{C2}$. However, as we demonstrate below, the terms with $r_{C1}$ and $
r_{C2}$ exactly cancel in the sum with the corresponding inelastic contribution, so their model
dependence is irrelevant for the finite result.

Table~\ref{tab:radii} presents our results for the effective charge radii $r_{C1}$ and $r_{C2}$,
for two models of the nuclear charge distribution. The ratios listed in the table are proven to be
independent of the nuclear model parameters $\lambda$ and $a$, thus making the corresponding
results valid for an arbitrary nucleus and both for the electronic and muonic atoms.

The formulas for $E^{(6)}_{\rm fns}$ have been derived in the nonrecoil limit, i.e., assuming the
infinite nuclear mass. This is different from the approach by Friar in Ref.~\cite{friar:79:ap}, in
which he replaced the lepton mass by the reduced mass of the system. We do not think such a
replacement is valid. However, recoil corrections for muonic atoms are significant and can be
partially accounted for by the $(\mu/m)^3$ scaling factor that comes from the square of the
nonrelativistic wave function at origin.

Equations (\ref{53})-(\ref{56}) can be compared with the analytical results by Friar derived within
a different approach \cite{friar:79:ap}. However, the formulas of Ref.~\cite{friar:79:ap} are so
complicated that a direct comparison is not possible, except for the state dependence which is in
perfect agreement. A comparison of numerical results presented in Sec.~\ref{sec:summary} shows a
reasonable but not perfect agreement. In order to verify our formulas, we performed a
high-precision numerical calculation, by solving the Dirac equation numerically and identifying the
$O(\alpha^2)$ finite nuclear size contribution, as described in Appendix \ref{app:2}. Perfect agreement
between analytical and numerical approaches confirms the correctness of Eqs.~(\ref{53})-(\ref{56}).

\begin{table}
  \caption{Various results for the exponential and the Gaussian models of the nuclear charge distributions.
\label{tab:radii}
}
\begin{ruledtabular}
\begin{tabular}{lcc}
\multicolumn{1}{l}{ } & \multicolumn{1}{c}{Exponential} & \multicolumn{1}{c}{Gaussian}
\\
\hline
\\[-5pt]
$\rho(q^2)$   & $\frac{\lambda^4}{(\lambda^2+q^2)^2}$ & $\exp\bigl(\frac{a\,q^2}{2}\bigr)$\\
$\rho(r)$     & $\frac{\lambda^3}{8\,\pi}\,e^{-\lambda\,r}$ &  $\frac1{(2\pi a)^{3/2}}\,\, {\exp\bigl(- \frac{r^2}{2\,a}\bigr)}$ \\
$r_C$         & $\frac{2\,\sqrt{3}}{\lambda}$ & $\sqrt{3\,a}$  \\
$V(r)$      & $\frac{1}{r} - \frac{e^{-\lambda\,r}}{r} - \frac{\lambda}{2}\,e^{-\lambda\,r}$
              & $\frac{1}{r}\,{\rm erf}\bigl(\frac{r}{\sqrt{2\,a}}\bigr)$ \\
$V^{(2)}(r)$ & $-\frac{r}{2} - \frac{2}{\lambda^2\,r} + \frac{1}{2\,\lambda}\,e^{-\lambda\,r} + \frac{2}{\lambda^2}\,\frac{e^{-\lambda\,r}}{r}$
              & $ -\sqrt{\frac{a}{2\,\pi}}\,\exp\bigl(-\frac{r^2}{2\,a}\bigr) - \frac{\bigl(a + r^2\bigr)}{2\,r}\,{\rm erf}\bigl(\frac{r}{\sqrt{2\,a}}\bigr)$ \\[2ex]
  $r_{C1}/r_C$ & $1.090\,044$ & $0.558\,872$ \\
  $r_{C2}/r_C$ & $1.068\,497$ & $1.014\,281$ \\
  $r_{CC}/r_C$ & $1.257\,433$ & $1.136\,219$ \\
  $r_{Z}/r_C$  & $1.558\,965$ & $1.514\,599$ \\
 \end{tabular}
\end{ruledtabular}
\end{table}

\section{Inelastic three-photon exchange correction in muonic deuterium}
The inelastic three-photon exchange nuclear structure correction has not yet been studied in the
literature and is the main topic of this work. With momenta of order $q\sim \sqrt{2\,m\,\Lambda}$,
which is much lower than the inverse of the nuclear size, the muon kinetic energy becomes
comparable to the characteristic nuclear excitation energy $\Lambda$, and thus the muon starts to
probe the nuclear structure and see individual nucleons. It means that the total correction does
not involve contributions coming from muon momenta of the order of the inverse of nuclear size, and
there is no place for the elastic high energy parts encoded in $r_{C1}$ and $r_{C2}$ effective
nuclear radii.

We represent the total nuclear structure correction $E^{(6)}$ as a sum of several parts,
\begin{equation}
  E^{(6)} = E^{(6)}_{1} + E^{(6)}_{2} + E_C + E^{(6)}_{\rm np} =
  E^{(6)}_{\rm fns} + E^{(6)}_{1, \rm pol} + E^{(6)}_{2, \rm pol} + E_C + E^{(6)}_{\rm np} \label{58}\,,
\end{equation}
where the elastic part $E^{(6)}_{\rm fns} = E^{(6)}_{1, \rm fns} + E^{(6)}_{2, \rm fns}$ was
calculated in the previous section, $E_{1}^{(6)}$ and $E_{1, \rm pol}^{(6)}$ are proportional to
$r_C^4$, $E_{2}^{(6)}$ and $E_{2, \rm pol}^{(6)}$ are proportional to $r_C^2$, $E_C$ is the known
Coulomb distortion correction \cite{pachucki:11:mud, hernandez:14}, and $E^{(6)}_{\rm np}$ is
the contribution due to the inelastic interaction with individual nucleons.

Below we calculate $E^{(6)}_{1}$ and $E^{(6)}_{2}$. Because the two-photon exchange nuclear structure
correction  $E^{(5)}_{\rm pol}$ was previously shown to be dominated by the electric dipole types
of the nuclear polarizability, we will assume that the same holds for the three-photon exchange
nuclear structure correction.

\subsection{Inelastic contribution $\bm \propto R^2$}
We represent the total $E_2^{(6)}$ correction as a sum of several parts
\begin{equation}
  E_2^{(6)} = E_{L2} + E_C + E_R + E_{H2}(p) \label{59}\,,
\end{equation}
calculated in the following.
Let us consider the two radiative photon exchange between the muon and the nucleus, taking into
account the Coulomb interaction $V$. The corresponding energy shift is
\begin{eqnarray}
\delta E &=&
i\,e^2\,\int\frac{d\,\omega}{2\,\pi}\,\int\frac{d^3k_1}{(2\,\pi)^3}\,\int\frac{d^3k_2}{(2\,\pi)^3}\,\omega^2\,
\frac{\Bigl(\delta^{ik}-\frac{k_1^i\,k_1^k}{\omega^2}\Bigr)}{\omega^2-k_1^2}\,
\frac{\Bigl(\delta^{jl}-\frac{k_2^j\,k_2^l}{\omega^2}\Bigr)}{\omega^2-k_2^2}
\nonumber \\ && \times\biggl[
\biggl\langle\bar\psi\biggl|\gamma^j\,e^{i\,\vec k_2\vec r}\,
\frac{1}{\not\!p-\gamma^0\,V - \gamma^o\,\omega-m}\gamma^i\,e^{i\,\vec k_1\vec r}\biggr|\psi\biggr\rangle
\nonumber \\ &&
+ \biggl\langle\bar\psi\biggl|\gamma^i\,e^{i\,\vec k_1\vec r}\,
\frac{1}{\not\!p-\gamma^o\,V+\gamma^0\,\omega-m}\,\gamma^j\,e^{i\,\vec k_2\vec r}\,\biggr|\psi\biggr\rangle\biggr]
\nonumber \\ && \times
\biggl\langle\phi_N\biggl|d^{\,k}\,\frac{1}{E_N-H_N-\omega}\,
d^{\,l}\biggr|\phi_N\biggr\rangle, \label{60}
\end{eqnarray}
where $\vec d$ is the dipole moment operator divided by the elementary charge. In the case of
deuteron $\vec d$ is equal to the position of the proton with respect to the mass center $\vec d =
\vec R$. In the nonrelativistic limit, $\delta E$ takes the well known form of Eq. (\ref{15}).
The corresponding low-energy $\alpha^6$ contribution is obtained from Eq.~(\ref{60}) by assuming
that the muon momenta are of the order of $m\,\alpha$. Then $E_0-H_0$  can be neglected in
comparison to $E_N-H_N$ and one obtains (with $d=3-2\,\epsilon$),
\begin{equation}
\delta_L E = \alpha^2\,\biggl\langle\phi\biggl|\frac{1}{r^4}\biggr|\phi\biggr\rangle_\epsilon\,
\frac{1}{d}\,\biggl\langle\phi_N\biggl| \vec R\,\,\frac{1}{E_N-H_N}\,\vec R\,\biggr|\phi_N\biggr\rangle
\label{61}\,,
\end{equation}
and
\begin{equation}
  \biggl\langle\phi\biggl|\frac{1}{r^4}\biggr|\phi\biggr\rangle_\epsilon
  = \bigl\langle\bigl[\nabla{\cal V}(r)\bigr]^2\bigr\rangle
  = \biggl\langle\frac{1}{r^4}\biggr\rangle +\phi^2(0)\,4\,\pi\,\biggl(-\frac{1}{2\,\epsilon} +2\,\ln\alpha + 2\biggr)
  \label{62}\,,
\end{equation}
were $\langle 1/r^4 \rangle$ is defined as an integral from a small radius $a$ to infinity with the
$1/a$ and $\ln a+\gamma$ terms subtracted out. The high-energy $\alpha^6$ part is obtained by
assuming that muon momenta are of the order $\sqrt{2\,m\,\Lambda}$. Then we can use the
explicit Coulomb correction
\begin{eqnarray}
\delta_H E &=&
-i\,e^2\,\int\frac{d\,\omega}{2\,\pi}\,\int\frac{d^3k_1}{(2\,\pi)^3}\,\int\frac{d^3k_2}{(2\,\pi)^3}\,\omega^2\,
\frac{\Bigl(\delta^{ik}-\frac{k_1^i\,k_1^k}{\omega^2}\Bigr)}{\omega^2-k_1^2}\,
\frac{\Bigl(\delta^{jl}-\frac{k_2^j\,k_2^l}{\omega^2}\Bigr)}{\omega^2-k_2^2}\,
\frac{4\,\pi\,\alpha}{(\vec k_1+\vec k_2)^2}
\nonumber \\ && {\rm Tr}\biggl[\biggr(
\gamma^j\,\frac{1}{\not\!p + \not\!k_2-m}\,\gamma^0\,\frac{1}{\not\!p - \not\!k_1-m}\,\gamma^i
+\gamma^0\,\frac{1}{\not\!p -\not\!k_1-\not\!k_2-m}\,\gamma^j\,\frac{1}{\not\!p - \not\!k_1-m}\,\gamma^i
\nonumber \\ &&
+\gamma^j\,\frac{1}{\not\!p +\not\!k_2-m}\,\gamma^i\,\frac{1}{\not\!p + \not\!k_1+\not\!k_2 -m}\gamma^0
\biggr)\,\,\frac{(\gamma^0+I)}{4}\biggr]\,\phi^2(0)
\nonumber \\ && \times
\biggl[\biggl\langle\phi_N\biggl|R^{\,k}\,\frac{1}{E_N-H_N-\omega}\,
R^{\,l}\biggr|\phi_N\biggr\rangle +
\biggl\langle\phi_N\biggl|R^{\,l}\,\frac{1}{E_N-H_N+\omega}\,
R^{\,k}\biggr|\phi_N\biggr\rangle\biggr] \label{63}\,,
\end{eqnarray}
where $k_1 = (\omega, \vec k_1)$, $k_2 = (-\omega, \vec k_1)$, and $p = (m,\vec 0)$. Assuming that
$(H_N-E_N)/m$ is small, one performs an expansion and obtains
\begin{eqnarray}
\delta_H E &=&
-\pi\,\alpha^3\,\phi^2(0)\,\frac{1}{d}\,\biggl[\biggl\langle\phi_N\biggl|\vec R
\frac{4}{H_N-E_N}\,
  \biggl(\frac{1}{2\,\epsilon}-1+\ln2-\ln\frac{(H_N-E_N)}{m}\biggr)\,\vec R\,\biggr|\phi_N\biggr\rangle\nonumber \\&&
  +\biggl\langle\phi_N\biggl|\vec R\biggl(\frac{1}{2\,\epsilon}+\frac{13}{2}-13\,\ln2
    -5\,\ln\frac{(H_N-E_N)}{m}\biggr)\vec R\,\biggr|\phi_N\biggr\rangle\biggr] \label{64}\,.
\end{eqnarray}
The sum of $\delta_L E$ in Eq. (\ref{61}) and $\delta_H E$ in Eq. (\ref{64}) gives the leading
Coulomb distortion correction $E_C$,
\begin{eqnarray}
  E_C &=& -\frac{1}{3}\,\biggl\langle\frac{1}{r^4}\biggr\rangle\,
    \biggl\langle\phi_N\biggl|\vec R\,\frac{1}{H_N-E_N}\,\vec R\,\biggr|\phi_N\biggr\rangle\nonumber \\ &&
    - \frac{4\,\pi}{3}\,\phi^2(0)\,\biggl\langle\phi_N\biggl|\,
    \vec R\,\frac{1}{H_N-E_N}\,\biggl[1+\ln\biggl(\frac{2\,m\,\alpha^2}{H_N-E_N}\biggr)\biggr]\vec R\,
    \biggr|\phi_N\biggr\rangle \label{65}\,,
\end{eqnarray}
where
\begin{eqnarray}
  \biggl\langle\frac{1}{r^4}\biggr\rangle_{nS} &=& \frac{8}{n^3}\,
  \biggl[-\frac{5}{3} + \frac{1}{2\,n} +\frac{1}{6\,n^2}+\gamma+\Psi(n)-\ln\frac{n}{2}\biggr]\label{66}\,,\\
  \biggl\langle\frac{1}{r^4}\biggr\rangle_{nP} &=& \frac{2\,(3\,n^2-2)}{15\,n^5}\label{67}\,,
\end{eqnarray}
and the relativistic correction $E_R$,
\begin{equation}
    E_R =-\pi\,\alpha^3\,\phi^2(0)\,\frac{r^2_s}{6}\,
    \biggl(\frac{1}{\epsilon}+\frac{41}{3}-26\,\ln2
    -10\,\ln\frac{\langle E \rangle_2}{m}\biggr) \label{68}\,,
\end{equation}
where $r^2_s = \langle R^{\,2}\rangle$ is the deuteron structure radius and
\begin{equation}
  \ln\frac{\langle E\rangle_2}{m} = \frac{1}{r_s^2}\,
  \biggl\langle\phi\biggl| \vec R\,\ln\frac{(H_N-E_N)}{m}\,\vec R\,\biggr|\phi\biggr\rangle \label{69}\,.
\end{equation}
Although there is no elastic high-energy part, the individual proton contributes
\begin{equation}
  E_{H2}(p) = -\pi\,\alpha^3\,\phi^2(0)\,\frac{r_p^2}{6}\,
  \biggl[\frac{1}{\epsilon} -2+4\,\gamma + 4\,\ln(r_{p2}\,m)\biggr]\,. \label{70}
\end{equation}
The last contribution $E_{L2}$ is exactly the same as the one of Eq.~(\ref{33}) with the deuteron
radii.

Finally, the total nuclear structure contribution $E^{(6)}_2$ $\propto R^2$ is given by the sum of
the elastic and inelastic parts,
\begin{eqnarray}
  E^{(6)}_2(nS) &=& E_C(nS)-\alpha^6\,m^3\,\frac{2}{3\,n^3}\,\biggl[
  r_d^2\,\biggl(\frac{9}{4n^2} - 3 -\frac{1}{n} +2\,\gamma -\ln\frac{n}{2}+\Psi(n)
  +\ln\alpha\biggr)\label{e2} \nonumber \\ &&
+r^2_s\, \biggl(\frac{47}{12}-\frac{13}{2}\,\ln2 -\frac{5}{2}\,\ln\frac{\langle E \rangle_2}{m}-\gamma\biggr)
+r_p^2\,\ln(r_{p2}\,m)\biggr]\,.
\end{eqnarray}
It is remarkable that the part of the elastic contribution depending on the effective deuteron
radius $r_{d2}$ does not show up in total $E^{(6)}_2$. Separately, the expression
for the inelastic contribution is
\begin{eqnarray}
E^{(6)}_{\rm 2,pol} = E^{(6)}_2 - E^{(6)}_{\rm 2,fns} - E_C &=& -\frac{2\,\alpha^6\,m^3}{3\,n^3}\,\delta_{l0}\,\biggl[
r^2_s\, \biggl(\frac{47}{12}-\frac{13}{2}\,\ln2 -\frac{5}{2}\,\ln\frac{\langle E \rangle_2}{m}\biggr) \label{71}\\ &&
+r_p^2\,\bigl(\gamma + \ln(r_{p2}\,m)\bigr)
-r_d^2\,\bigl(\gamma+\ln(r_{d2}\,m)\bigr)\biggr] \nonumber\,.
\end{eqnarray}
The averaged excitation energy $\langle E\rangle_2$ has been calculated using the AV18
potential \cite{wiringa:95} with the result
\begin{equation}
  \langle E\rangle_2 = 7.374\;{\rm MeV},
\end{equation}
exactly the same as for $\mu$D and $e$D, because the lepton mass cancels out between the left and
the right side of Eq. (\ref{69}).

\subsection{Inelastic contribution $\bm \propto R^4$}
We represent the total $E_1^{(6)}$ correction as a sum of four parts
\begin{equation}
  E_1^{(6)} = E_{L1} + E_Q + E'_Q + E_{H1}(p) \label{72}\,.
\end{equation}
The middle energy contribution $E_Q$ comes from momenta $q\sim \sqrt{2\,m\,\Lambda}$. We derive it by considering
the nonrelativistic three Coulomb photon exchange,
\begin{equation}
  E_Q = -\biggl\langle\phi,\phi_N\biggl|
  \frac{\alpha}{|\vec r-\vec R|}\,\frac{1}{E-H_0 + E_N - H_N}\,
  \frac{\alpha}{|\vec r-\vec R|}\,\frac{1}{E-H_0 + E_N - H_N}\,
  \frac{\alpha}{|\vec r-\vec R|}\biggr|\phi,\phi_N\biggr\rangle\label{73}\,,
\end{equation}
where $H_0$ is the lepton kinetic energy operator. In the corresponding electronic matrix element
$P_Q$, one neglects the lepton binding energy $E$,
\begin{equation}
P_Q  =
-\phi^2(0)\,\alpha^3\,4\,m^2\,(4\,\pi)^3\int\frac{d^dq_1}{(2\,\pi)^d}\,\frac{d^dq_2}{(2\,\pi)^d}\,
\frac{e^{i\,\vec q\vec R_1}}{q^2}\,\frac{1}{q^2+2\,m\,\Lambda}\,
\frac{e^{-i\,(\vec q+\vec q')\vec R_2}}{(\vec q + \vec q')^2}\,\frac{1}{q'^2+2\,m\,\Lambda'}\,
\frac{e^{i\,\vec q'\vec R_3}}{q'^2} \label{74}\,,
\end{equation}
expands in $R_i$, keeping terms $\propto R^4$, with the result
\begin{equation}
  E_Q = -\phi^2(0)\,\alpha^3\,m^2\,\pi\,\biggl[
  \langle R^4\rangle\,\biggl(\frac{4}{15}\,\ln2-\frac{2}{5}\biggr)
  +\langle R^2\rangle^2\biggl(-\frac{10}{27} + \frac{2}{3}\,\ln2 +\frac{2}{9}\,
  \ln\frac{\langle E \rangle_1}{m} -\frac{8}{9}\,\beta - \frac{1}{9\,\epsilon}\biggr)\biggr]\label{75}\,,
\end{equation}
where
\begin{eqnarray}
  \ln\frac{\langle E\rangle_1}{m}&=&
  -\frac{1}{\langle R^2\rangle^2}\,\biggl[
    \langle 0|R^2\,\ln\frac{(H-E)'}{m}\,R^2|0\rangle
    -\frac{6}{5}\,\langle 0|R^i\,\ln\frac{(H-E)'}{m}\,R^2\,R^j|0\rangle
  \nonumber \\ &&
  +\frac{3}{10}\,\langle 0|(R^i\,R^j-\delta^{ij}\,R^2/3)\,\ln\frac{(H-E)'}{m}\,(R^i\,R^j-\delta^{ij}\,R^2/3)|0\rangle
  \biggr] \label{76}\,.
\end{eqnarray}
The average energy $\langle E\rangle_1$ does not depend on the lepton mass $m$, since the
dependence on $m$  cancels out between the left and right side of above equation. We calculate
$\langle E\rangle_1$ by using the AV18 deuteron potential \cite{wiringa:95}, with the result
\begin{equation}
\langle E\rangle_1 = 2.932\;{\rm MeV}\,.\label{77}
\end{equation}
Equation~(\ref{75}) involves the dimensionless parameter $\beta$ defined by
\begin{eqnarray}
  \beta =
  -\sum_{\Lambda_1,\Lambda_2}{\!\!\!}'\,
  &\Bigl[&
  \langle 0|R^i\,R^j+3\,\delta^{ij}\,R^2|\Lambda_1\rangle\,\langle\Lambda_1|R^i|\Lambda_2\rangle\,
  \langle\Lambda_2|R^j|0\rangle
  \nonumber \\[-2ex] &+&
  \langle 0|R^i|\Lambda_1\rangle\,\langle\Lambda_1|\delta^{ij}\,R^2-3\,R^i\,R^j|\Lambda_2\rangle\,
  \langle\Lambda_2|R^j|0\rangle
  \nonumber \\ &+&
  \langle 0|R^i|\Lambda_1\rangle\,\langle\Lambda_1|R^j|\Lambda_2\rangle\,
  \langle\Lambda_2|\delta^{ij}\,R^2-3\,R^i\,R^j|0\rangle\Bigr]\,\frac{3}{10\,\langle R^2\rangle^2}\,
  f\biggl(\frac{\Lambda_1}{\Lambda_2}\biggr) \label{78}\,,
\end{eqnarray}
with
\begin{equation}
f(x)  = x\,\ln\biggl(1+\frac{1}{\sqrt{x}}\biggr) - \sqrt{x} - \ln(1+\sqrt{x}) \label{79}\,.
\end{equation}
Since $f$ weakly depends on its argument, one can replace the argument of $f$ in Eq.~(\ref{78}) by
its averaged value to obtain
\begin{equation}
  \beta = f\biggl(\biggl\langle\frac{\Lambda_1}{\Lambda_2}\biggr\rangle\biggr) \label{80}\,.
  \end{equation}
For the estimation of $\beta$ we will assume that $\langle\Lambda_1/\Lambda_2\rangle = 1, 2, 1/2$
and thus obtain $\beta = -1.0\,(0.2)$.

There is an additional contribution $E'_Q$ that includes the finite proton size. It is obtained by
inserting the proton electric formfactor in the Coulomb interaction in Eq. (\ref{73}) and expanding
in $R_i$ up to the second order,
\begin{equation}
  E'_Q =  \phi^2(0)\,\alpha^3\,m^2\,r_s^2\,r^2_p\,\frac{4\,\pi}{9}\,\biggl(\frac{13}{3} + \frac{1}{2\,\epsilon}
     -5\,\ln 2 -\frac{1}{r_s^2}\,\langle 0|\vec R\,\ln\frac{(H-E)}{m}\,\vec R|0\rangle\biggr)\label{81}\,.
\end{equation}
The remaining contributions $E_{L1}$ is given by Eq. (\ref{32}) with the deuteron radii, whereas
$E_{H1}(p)$ is given by Eq. (\ref{48}) with the proton charge radii.

Adding all parts together, the total nuclear structure contribution $E^{(6)}_1$ $\propto R^4$  is
\begin{eqnarray}
 E^{(6)}_1(nS) &=& \alpha^6\,m^5\,\frac{4}{9\,n^3}\,\biggl[r_d^4\,\biggl(-\frac{1}{n} + \gamma -\ln\frac{n}{2}+\Psi(n)
        + \ln\alpha\biggr) + r_{dd}^4\,\frac{3}{20\,n^2}  \label{e1}\\ &&
  -r_{ss}^4\biggl(\frac{3}{5}\,\ln2-\frac{9}{10}\biggr)
  +r_s^4\,\biggl(\frac{1}{3} - \frac{3}{2}\,\ln2 - \frac{1}{2}\,\ln\frac{\langle E \rangle_1}{m} + 2\,\beta\biggr) \nonumber \\
    &&+ r_p^4\,\biggl(2+\gamma +\ln(r_{p1}\,m)\biggr)
        + r_s^2\,r^2_p\,\biggl(\frac{10}{3}-5\,\ln 2 - \ln\frac{\langle E\rangle_2}{m}\biggr)\biggr] \nonumber\,.
\end{eqnarray}
Again, the part of the elastic contribution depending on the effective deuteron radius $r_{d1}$
is not present in  total $E^{(6)}_1$. The expression for the separate  inelastic contribution is
\begin{eqnarray}
E^{(6)}_{\rm 1,pol} &=&   E^{(6)}_1 - E^{(6)}_{\rm 1,fns} = \frac{\alpha^6}{n^3}\,m^5\,\delta_{l0}\,\biggl[
  -r_{ss}^4\,\biggl(\frac{4}{15}\,\ln2-\frac{2}{5}\biggr)
  +\frac{2}{9}\,r_s^4\,\biggl(\frac{2}{3} - 3\,\ln2 - \ln\frac{\langle E \rangle_1}{m} + 4\,\beta\biggr) \nonumber \\
    &&\hspace*{-7ex} + \frac{4}{9}\,r_p^4\,
    \biggl(2+\gamma +\ln(r_{p1}\,m)\biggr)
        + \frac{4}{9}\,r_s^2\,r^2_p\,\biggl(\frac{10}{3}-5\,\ln 2 - \ln\frac{\langle E\rangle_2}{m}\biggr)
        -\frac{4}{9}\,r_d^4\, \biggl(2+\gamma +\ln(r_{d1}\,m)\biggr)\biggr].\nonumber \\ \label{82}
\end{eqnarray}

\subsection{Total inelastic part}

The sum $E^{(6)}_{\rm 1,pol} + E^{(6)}_{\rm 2,pol}$, as given by Eqs.~(\ref{82}) and (\ref{71}), is
the total three-photon exchange inelastic nuclear structure contribution, which is the main result
of this work. It should be pointed out that several approximations have been made in our derivation
of this result. First, we ignored the magnetic dipole and the electric quadrupole moments of
deuteron. Second, we neglected the higher orders in $(H_N-E_N)/m$. These approximations contribute
to the uncertainty of the inelastic part, which we estimate as 10\%.

The remaining part of the total three-photon exchange nuclear structure contribution of
Eq.~(\ref{58}) is the contribution due to the interaction with individual nucleons $E^{(6)}_{\rm np}$.
We have little knowledge about the inelastic three-photon exchange between the muon
and the proton but we expect it could be accounted for in terms of the same effective radii
$r_{p1}$ and $r_{p2}$ as in the elastic part. We estimate the uncertainty associated with
$E^{(6)}_{\rm np}$ in $\mu$H and in $\mu$D by applying Eq.~(\ref{53}) to the proton ($r_C \to
r_p$) and making the following substitution,
\begin{eqnarray}
  \ln r_{p1} &\rightarrow&  \ln r_{p1}\,\pm 1\,, \nonumber \\
  \ln r_{p2} &\rightarrow&  \ln r_{p2}\,\pm 1\,.
\end{eqnarray}
It should be mentioned that $E^{(6)}_{\rm np}$ does not contribute to the $\mu$D-$\mu$H
isotope shift difference.

\section{Inelastic three-photon exchange correction in ordinary deuterium}
The total inelastic nuclear structure $\alpha^6$ correction for ordinary deuterium is split into
three parts,
\begin{equation}
  E^{(6)} = E_{L2} + E_R + E_{H2}(p) + E^{(6)}_{\rm np}\,, \label{83}
\end{equation}
where $E_{L2}$ is given by Eq.~(\ref{33}) with the deuteron radii, and $E_{H2}(p)$ by Eq.
(\ref{52}) with the proton radii, while $E_R$ is a Coulomb correction to the electric dipole
polarizability, as given by Eq.~(\ref{63}), and $E^{(6)}_{\rm np}$ is the correction due to
the interaction with individual nucleons.

Assuming that $H_N-E_N$ is much larger than the electron mass $m$, we obtain
\begin{equation}
E_R =
-\pi\,\alpha^3\,\phi^2(0)\,\frac{1}{d}\,
\biggl\langle\phi_N\biggl|R^{\,k}\biggl[
\,\frac{1}{2\,\epsilon} + \frac{5}{2} - 2\,\ln\frac{2\,(H_N-E_N)}{m}\biggr]
\,R^{\,k}\biggr|\phi_N\biggr\rangle + O\biggl(\frac{m}{H_N-E_N}\biggr) \label{84}\,.
\end{equation}
We note that the neglected $O(m/(H_N-E_N))$ terms do not vanish for the $l>0$ states. Therefore,
the nuclear polarizability correction does not vanish for the $l>0$ states, but it is additionally
suppressed by the ratio of the electron mass to the nuclear excitation energy.

The correction due to the interaction with individual nucleons $E^{(6)}_{\rm nucleon}$ is expected
to be small and, moreover, it cancels out in the $e$D-$e$H isotope shift difference. In order to
estimate the three-photon exchange of the bound electron with the proton, we use the same
argumentation as in Ref.~\cite{khriplovitch:00} to obtain
\begin{equation} \label{84a}
  E^{(6)}_{\rm np}(p) = \frac{2\,\pi}{3}\,\alpha^3\,\phi^2(0)\,r_p^2\,\ln\frac{\bar E_p }{m}\,,
\end{equation}
and assume the uncertainty of 100\%. The above correction is proportional to the squared charge
radius, so the corresponding contribution for the neutron is negligible.

Our final result for the three-photon exchange nuclear structure correction in deuterium is given
by
\begin{eqnarray}
  E^{(6)}(nS) &=& -\alpha^6\,m^3\,\frac{2}{3\,n^3}\,\biggl[r_d^2\,\biggl(\frac{9}{4n^2} - 3 -\frac{1}{n} + \gamma -\ln\frac{n}{2}+\Psi(n)
    +\ln \alpha\biggr)\label{eD}\\&&\hspace*{13ex}
   +r_s^2\,\biggl(\frac{23}{12} - \ln\frac{2\,\langle E \rangle_2}{m}\biggr)
   +r_p^2\,\bigl(\gamma + \ln(r_{p2}\,m)\bigr)\biggr] + E^{(6)}_{\rm np}\,.\nonumber
\end{eqnarray}
Separately, the inelastic part $E^{(6)}_{\rm pol} = E^{(6)} - E^{(6)}_{\rm fns}$ is
\begin{equation}
  E^{(6)}_{\rm pol} = -\frac{2\,\alpha^6}{3\,n^3}\,m^3\,\delta_{l0}\,\biggl[
    r_s^2\,\biggl(\frac{23}{12} - \ln\frac{2\,\langle E \rangle_2}{m}\biggr)
   +r_p^2\,\bigl(\gamma + \ln(r_{p2}\,m)\bigr)
   -r_d^2\,\bigl(\gamma + \ln(r_{d2}\,m)\bigr)\biggr]
   + E^{(6)}_{\rm np} \label{85}\,.
\end{equation}
We note that the fermion mass $m$ cancels exactly in the expression in square brackets in the above equation.

\section{Results and summary}
\label{sec:summary}

Our numerical results for the three-photon exchange nuclear structure corrections are presented in
Table~\ref{tab:results}. The elastic part has been calculated with the exponential model of the
nuclear charge distribution. It is displayed in the table separately for the comparison with the
literature results. This part does not bear any uncertainty because its dependence on the charge
distribution model cancels out exactly in the sum with the inelastic part. We observe a reasonable
(although not perfect) agreement with the literature results summarized in Table~\ref{tab:results}.

The inelastic three-photon exchange nuclear structure correction was calculated only for the
electronic and muonic deuterium atoms; the corresponding results are presented in
Table~\ref{tab:results}. We find that the inelastic contribution for deuterium is of opposite sign
as compared to its elastic counterpart and changes significantly the total $m\alpha^6$ nuclear
structure contribution. In the case of $e$D, the change is of about 30\%, while for $\mu$D, the
inelastic part reverses the sign of the overall contribution. For electronic and muonic hydrogen, we
present only estimations for the inelastic three-photon exchange contribution.

Our results for the three-photon exchange nuclear structure corrections affect determinations of
the hydrogen-deuterium nuclear charge radii differences derived from the spectroscopic observations
of the isotope shifts in electronic and muonic hydrogen and deuterium
\cite{parthey:10,jentschura:11,pohl:16}. For the electronic H-D isotope shift of the $1S$--$2S$
transition, our result shifts the total theoretical prediction by 0.8~kHz, which is slightly larger
than the theoretical error of 0.6~kHz assumed in Ref.~\cite{jentschura:11}. There are, however,
further corrections to the summary of theoretical contributions presented in
Ref.~\cite{jentschura:11}, so we had to update it. Our review of the present status of the theory of
the H-D isotope shift described in Appendix~\ref{app:HD} leads us to the updated result for the
nuclear charge radius difference determined from the measurement of the H-D isotope shift of the
$1S$--$2S$ transition \cite{parthey:10},
\begin{equation}
\delta r^2 [\mbox{\rm electronic}] \equiv r_d^2-r_p^2\, = 3.820\,70\,(31)\, {\rm fm}^2\,,
\end{equation}
which agrees with but is twice as accurate as the previous value of $3.820\,07\,(65)$~fm$^2$
obtained in Ref.~\cite{jentschura:11}.

For muonic hydrogen and deuterium, our result for the inelastic three-photon exchange nuclear
structure contribution to the $2P_{1/2}$--$2S$ transition energy of $0.008\,75\,(88)$~meV shifts
the deuteron-proton charge radius difference determined in Ref.~\cite{pohl:16} by $0.0014$~fm$^2$,
with the result
\begin{equation}
\delta r^2 [\mbox{\rm muonic}] \equiv r^2_d-r^2_p\, = 3.8126\,(34)\, \mbox{\rm fm}^2\,.
\end{equation}
The results derived from the electronic and muonic atoms disagree by about $2\,\sigma$, which
confirms the discrepancy previously observed in Ref.~\cite{pohl:16}.

\begin{table}[!htb]
\caption{\label{tab:results}
Numerical results for the three-photon exchange
nuclear structure corrections. Numerical values include the leading recoil effect by the multiplicative reduced-mass
prefactor $(\mu/m)^3$.
Elastic contributions are obtained with the exponential parametrization of the nuclear charge distribution, with the
following values of nuclear radii: $r_p = 0.84087$~fm, $
r_d = 2.12562$~fm, $r_C(^3{\rm He}) \equiv r_h = 1.973$~fm \cite{sick:14}, $r_C(^4{\rm He}) \equiv
r_{\alpha} = 1.681$~fm \cite{sick:08}.}
\begin{ruledtabular}
\begin{tabular}{r r c c c l}
transition & units &\centt{Elastic}  & \centt{Inelastic} & \centt{Sum} & \centt{Elastic by others} \\
\hline\\[-7pt]
$E^{(6)}(2S{\rm-}1S,e{\rm{H}})$              & Hz  & $-584$       & $-344\,(344)$   &  $-928\,(344)$      & $-587\,(2)^{a}$ \\[1ex]
$E^{(6)}(2S{\rm-}1S,e{\rm D-}e{\rm H})$      & Hz  & $-2\,846$    &  $817\,(41)$    &  $-2\,029\,(41)$    & $-2\,834\,(13)^{a}$ \\[1ex]
$E^{(6)}(2P_{1/2}{\rm-}2S,\mu{\rm H})$       & meV & $-0.001\,27$ & $\pm 0.000\,27$ &  $-0.001\,27\,(27)$ & $-0.001\,34^{b}$    \\[1ex]
$E^{(6)}(2P_{1/2}{\rm-}2S,\mu{\rm D})$       & meV & $-0.006\,56$ & $0.008\,75\,(88)(27)^{\dag}$ & $\ \ \ 0.002\,19\,(88)(27)^{\dag}$     & $-0.006\,50\,(60)^{c}$  \\[1ex]
$E^{(6)}(2P_{1/2}{\rm-}2S,\mu ^3{\rm He}^+)$ & meV & $-0.384\,7$  & unknown         &                     & $-0.378\,6\,(60)^{d}$  \\[1ex]
$E^{(6)}(2P_{1/2}{\rm-}2S,\mu ^4{\rm He}^+)$ & meV & $-0.304\,8$  & unknown         &                     & $-0.311\,5\,(140)^{e}$     \\
\end{tabular}
\end{ruledtabular}
\flushleft\noindent
$^a$ CODATA \cite{mohr:16:codata}. \\
$^b$ Ref. \cite{antognini:13:ap}, the difference of entries ``Our choice'' and
``Non-rel. finite-size'' in Table~2 of that work,
$-0.0019\,r_p^2$\,.\\
$^c$ Ref. \cite{krauth:16:ap}, the sum of entries $r_3$ and $r_3'$ in
Table 2 of that work, $
-0.002\,124\,(4)\,r_d^2 + 0.003\,10\,(60)\ {\rm meV}$\,.\\
$^d$ Ref. \cite{franke:17}, the sum of entries $r_3$ and $r_3'$ in Table 2 of that work,
$-0.1288\,(13)\,r_h^2 + 0.1177\,(33)\ {\rm meV}$\,.\\
$^e$ Ref. \cite{diepold:18}, the sum of entries $r_3$ and $r_3'$ in Table 4 of that work,
$-0.1340\,(30)\,r_{\alpha}^2 + 0.0672\,(112)$ meV.\\
$^{\dag}$ the second uncertainty comes from the interaction with individual nucleons and cancels in the $\mu$D-$\mu$H isotope shift.
\end{table}

In summary, we have calculated the complete three-photon exchange $O(\alpha^2)$ nuclear structure
correction to energy levels and the isotope shift of hydrogen-like muonic and electronic atoms. Our
formula for the elastic contribution is valid for an arbitrary hydrogenic system and is much
simpler than corresponding formulas in the literature \cite{friar:79:ap}. The inelastic part has
been derived for muonic and electronic deuterium  only. Calculations of the three-photon inelastic
contribution for He$^+$ and heavier elements are possible but are more complicated. At the same time,
one may expect the inelastic contribution to be as large as the elastic part, which is a sizeable
correction in He$^+$, about 1\% of the total nuclear nuclear size effect.

\acknowledgments
The authors acknowledge support from the National Science Center of Poland
(Grant No. 2012/04/A/ST2/00105).
V.P. acknowledges support also from the Czech Science Foundation - GA\v{C}R (Grant No. P209/18-00918S), and
V.A.Y. acknowledges support also from the Ministry of Education and Science of the Russian Federation
Grant No.~3.5397.2017/6.7.


\appendix

\section{Dimensional regularization for bound states}
\label{app:1} The principles of dimensional regularization for bound states have been described in
Ref. \cite{pachucki:06:hesinglet}. Here we only present formulas without derivation which have been
used in the presented calculations. The dimension of space is assumed to be $d=3-2\,\epsilon$. The
surface area of the $d$-dimensional unit sphere is
\begin{equation}
\Omega_d = \frac{2\,\pi^{d/2}}{\Gamma(d/2)}\,.\label{a1}
\end{equation}
The Coulomb potential in $d$ dimensions is of the form
\begin{equation}
{\cal V}(r) = \int\frac{d^d k}{(2\,\pi)^d}\,\frac{4\,\pi}{k^2}\,e^{i\,\vec k\cdot\vec r} = \frac{C_1}{r^{1-2\,\epsilon}}\,, \label{a2}
\end{equation}
where
\begin{equation}
C_1 = \pi^{\epsilon-1/2}\,\Gamma(1/2-\epsilon) \label{a3}\,.
\end{equation}
The elastic contribution involves another potential of the form
\begin{equation}
{\cal V}^{(2)}(r) = \int \frac{d^d k}{(2\,\pi)^d}\,\frac{4\,\pi}{k^4}\, e^{i\,\vec k\cdot\vec r}
= C_2\,r^{1+2\,\epsilon},\label{a4}
\end{equation}
where
\begin{equation}
C_2 = \frac{1}{4}\,\pi^{\epsilon-1/2}\,\Gamma(-1/2-\epsilon).
\label{a5}
\end{equation}
Futher, we used the following integration formulas
\begin{eqnarray}
  \int_\Lambda d^d r\,\bigl[{\cal V}(r)\bigr]^3 &=&
 -[(4\,\pi)^\epsilon\,\Gamma(1+\epsilon)]^2\,4\,\pi\,\,\biggl[\frac{1}{4\,\epsilon} + \frac{1}{2}+\gamma+\ln(\Lambda)\biggr]\,,\label{a6}\\
\int\frac{d^dk}{(2\pi)^d}\frac{1}{k^{2\alpha}}\frac{1}{(k-q)^{2\beta}}&=&\frac{[q^2]^{\frac{d}{2}-\alpha-\beta}}{[4\pi]^{\frac{d}{2}}}
\frac{\Gamma(\alpha+\beta-\frac{d}{2})\,\Gamma(\frac{d}{2}-\alpha)\,\Gamma(\frac{d}{2}-\beta)}{\Gamma(d-\alpha-\beta)\,\Gamma(\alpha)\,\Gamma(\beta)}
\label{a7}\,,
\end{eqnarray}
and \cite{davydychev:93}
\begin{eqnarray}
  I &=&
  \int \frac{d^d k}{(2\,\pi)^d}\, \int \frac{d^d q}{(2\,\pi)^d}\,
  \frac{1}{[k^2]^{n_1}}\,\frac{1}{[(k-q)^2+m_2^2]^{n_2}}\,\frac{1}{[q^2+m_3^2]^{n_3}}
  \nonumber \\ &=&
  \frac{m_3^{2\,(d-n_1-n_2-n_3)}}{(4\,\pi)^d}\,
  \frac{\Gamma(d/2-n_1)\, \Gamma(n_1+n_2-d/2)\, \Gamma(n_1+n_3-d/2)\, \Gamma(n_1+n_2+n_3-d)}
       {\Gamma(2\,n_1+n_2+n_3-d)\, \Gamma(n_2)\, \Gamma(n_3)\, \Gamma(d/2)}\nonumber \\&&
\times       _2F_1(n_1+n_2+n_3-d,n_1+n_2-d/2,2\,n_1+n_2+n_3-d,1- m_2^2/m_3^2)\,.\label{a8}
\end{eqnarray}

\section{Numerical verification of the elastic contribution}
\label{app:2} The finite nuclear size (fns) correction can be calculated numerically to all orders
in $\Za$, by computing the energy eigenvalue of the Dirac equation with the extended-size nuclear
potential and subtracting the analytical point-nucleus result. Knowing the leading $\alpha^4$ and
$\alpha^5$ fns corrections analytically, we also can identify the higher-order fns residual from
the numerical all-order results.

The main problem in determining the fns correction numerically is that the corresponding effect is
very small for light electronic atoms. So, for the $2s$ state of hydrogen, the relativistic
$O(\alpha^2)$ fns correction yields a $1\times 10^{-13}$ fraction of the binding energy. In order
to  make an extensive comparison between the numerical and analytical approaches, we performed
numerical calculations for $Z$ as low as $Z = 0.25$. To make sure that possible numerical
uncertainties do not interfere with the comparison, we determined the binding energies with a
20-digit numerical precision and made sure that the nuclear charge distribution model is exactly
the same as in analytical calculations.

In order to compute the eigenvalues of the Dirac equation, we use the Dual Kinetic Balance method
\cite{shabaev:04:DKB} with the finite basis set of $B$-splines. Because of high accuracy demands,
we implemented this method in quadruple (about 32 digits) arithmetics, similarly as it was done
recently in calculations of the recoil corrections \cite{yerokhin:15:recprl}. About 200-250 basis
functions were sufficient to reach the required 20-digit numerical accuracy for the binding
energies.

We obtained the relativistic fns correction $E_{\rm fns}^{(6+)}$ that contains contributions of order
$(\Za)^6$ and higher as
\begin{align}
E_{\rm fns}^{(6+)} = E_{\rm fns} - E_{\rm fns}^{(4)} - E_{\rm fns}^{(5)}\,,\label{b1}
\end{align}
where $E_{\rm fns}$ is determined numerically by solving the Dirac equation, $E_{\rm fns}^{(4)}$ is
given by Eq.~(\ref{06}), $E_{\rm fns}^{(5)}$ is given by Eq.~(\ref{13}), and $r_Z$ is evaluated for
the same nuclear model as in the numerical calculation.

\begin{figure*}[!htb]
\centerline{
\resizebox{0.9\textwidth}{!}{%
  \includegraphics{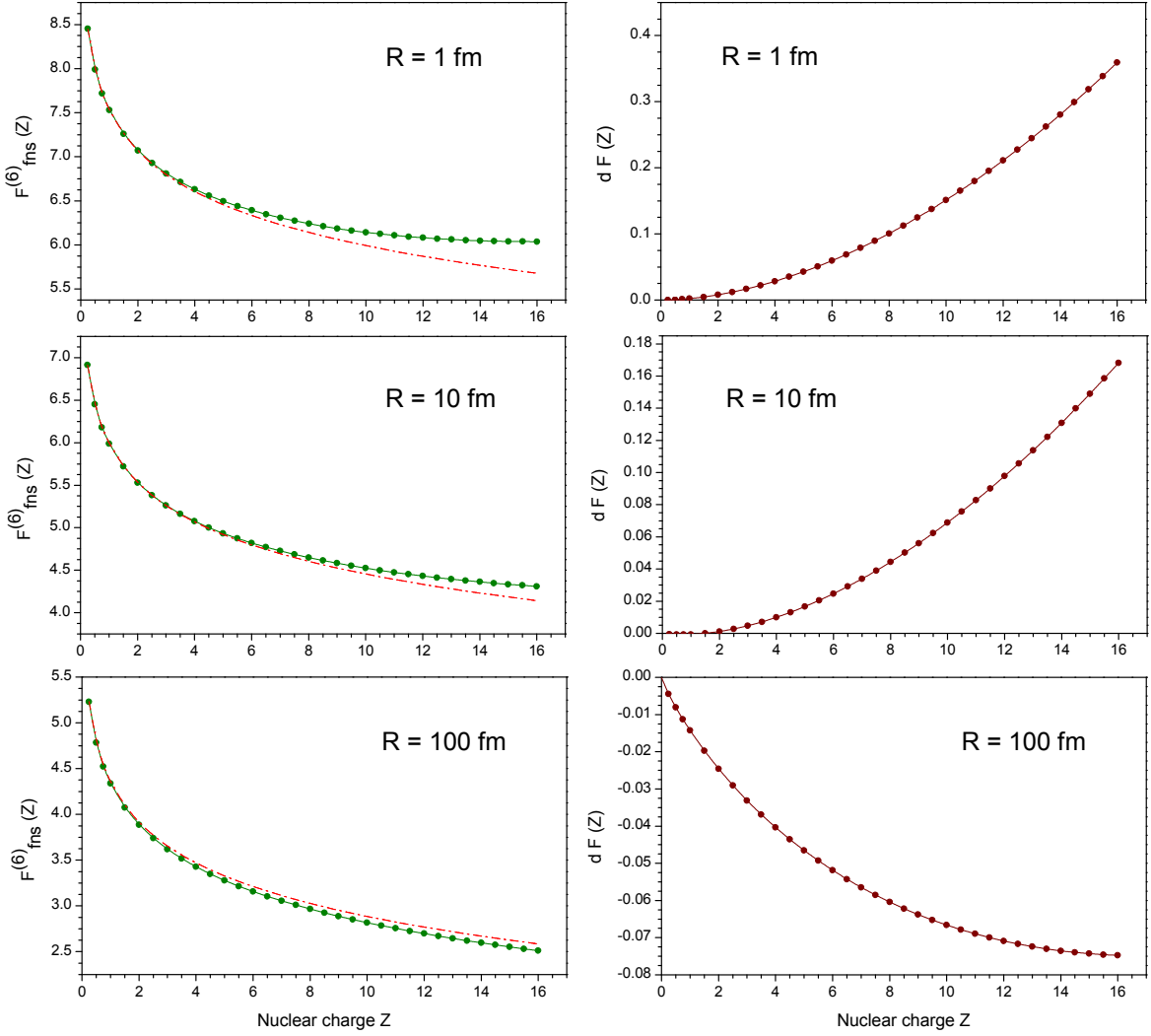}
}}
 \caption{The relativistic finite nuclear size correction for the hydrogenic $1s$ state is plotted  as a function of $Z$
 for three different nuclear radii: $r_C = 1$~fm (upper row),
 $r_C = 10$~fm (middle row), and $r_C = 100$~fm (lower row).
In each row, the left graph shows a comparison of the numerical function $F_{\rm fns}^{(6+)}$ (filled dots and solid line, green) with
the analytical function $F_{\rm fns}^{(6)}$ (dash-dotted line, red); the right graphs show the remainder function
$\delta F = F_{\rm fns}^{(6+)}-F_{\rm fns}^{(6)}$.
 \label{fig:1}}
\end{figure*}

The comparison of our all-order numerical results for $E_{\rm fns}^{(6+)}$ with the analytical
$(\Za)^6$ result $E^{(6)}$ given by Eq.~(\ref{53}) is presented in Fig.~\ref{fig:1}. Both numerical
and analytical results are obtained with the exponential model of the nuclear charge distribution
(see Table~\ref{tab:radii}). We plot the scaled function with the leading $\Za$, $r_C$, and $n$
dependence removed,
\begin{align}
F_{\rm fns}^{(6+)} = \frac{E_{\rm fns}^{(6+)}}{m^3 (Z\alpha)^6\,r_C^2/n^3}\,.\label{b2}
\end{align}
As can be seen from the figure, agreement between the numerical and analytical results is
excellent.

\section{Hydrogen-deuterium $\bm{1S}$--$\bm{2S}$ isotope shift}
\label{app:HD}

In this section we update the summary of all available theoretical contributions for the $e$H-$e$D
isotope shift of the $1S$--$2S$ transition frequency reviewed previously by Jentschura {\em et al.}
\cite{jentschura:11}. We use the following values for the fundamental constants, the fine-structure
constant,
$$
\alpha^{-1} = 137.035\,999\,139\,(31)\,,
$$
and the Rydberg constant,
$$
R_{\infty}c = 3.289\,841\,960\,355\,(19)\times 10^{15}\,{\mbox Hz}\,,
$$
from CODATA~2014 \cite{mohr:16:codata}. The electron-proton mass ratio we take from the recent
measurement by Hei{\ss}e {\em et al.} \cite{heisse:17},
$$
\frac{m_p}{m_e} = 1\,836.152\,673\,346\,(81)\,.
$$
We note that this value is twice as accurate but 3\,$\sigma$ off from the CODATA~2014 value
\cite{mohr:16:codata}. For the electron-deutron mass ratio we use the CODATA value
\cite{mohr:16:codata},
$$
\frac{m_D}{m_e} = 3\,670.482\,967\,85\,(13)\,.
$$

In the present section we follow the notations and conventions of Ref.~\cite{jentschura:11}. We
will not repeat the full review of the theory but only indicate the entries therein that need to be
updated. The changes are as follows.

{\em (i)} The updated result for the leading (Dirac) contribution to the isotope shift (Eq.~(28) of
Ref.~\cite{jentschura:11}) is
\begin{align}
\Delta f_{\rm i} = 671\,004\,071.107\,(64)~\mbox{\rm kHz}\,,
\end{align}
the change being due to the updated values of the electron-nucleus mass ratios.

{\em (ii)} Our present result for the two-photon exchange nuclear structure correction specified by
Eqs.~(\ref{E5H}) and (\ref{E5D}),
\begin{align}
E^{(5)}(\mbox{H--D},1S\mbox{\rm--}2S) = 19.12\,(20)~\mbox{\rm kHz}\,,
\end{align}
replaces the sum of $\Delta \nu_9$ given by Eq.~(40) of Ref.~\cite{jentschura:11} and $E_{\rm
NS,(b)}$ given by Eq.~(45) therein, amounting to $19.11\,(2)$~kHz.

{\em (iii)} Our present result for the three-photon exchange nuclear structure correction,
\begin{align}
E^{(6)}(\mbox{H--D},1S\mbox{\rm--}2S) = -2.029\,(41)~\mbox{\rm kHz}\,,
\end{align}
replaces the sum of $E_{\rm NS,(c)}$ given by Eq.~(47) of Ref.~\cite{jentschura:11} and $\Delta
\nu_{11}=\pm 0.5$~kHz given by Eq.~(43) therein, amounting to $-2.828 \pm 0.5$~kHz.

{\em (iv)} The entry for the higher-order pure recoil $\nu_5 = -3.41\,(32)$~kHz (Eq.~(33) of
Ref.~\cite{jentschura:11}) is replaced by the complete all-order (in $\Za$) result by Yerokhin and
Shabaev \cite{yerokhin:15:recprl,yerokhin:16:recoil}. The corresponding correction to the energy is
\begin{align}
\delta E(nS) = \frac{m^2}{M} \frac{(\Za)^5}{\pi n^3} \biggl[\Za \left(4 \ln 2 - \frac72\right)\pi  + (\Za)^2 \,G_{\rm rec}
+ \delta_{\rm fns} P \biggr]\,,
\end{align}
where $G_{\rm rec}(1S,Z=1) = 9.720\,(3)$, $G_{\rm rec}(2S,Z=1) = 14.899\,(3)$, $\delta_{\rm fns}
P(nS,{\rm H}) = -0.000\,184\,(1)$ in the case of hydrogen and $\delta_{\rm fns} P(nS,{\rm D}) =
-0.000\,786\,(6)$ for deuteron \cite{yerokhin:15:recprl,yerokhin:16:recoil}. In the result, the
updated contribution is
\begin{align}
\Delta \nu_{5} = -3.058~\mbox{\rm kHz}\,.
\end{align}

{\em (v)} For the radiative recoil contribution (Eq.~(36) of Ref.~\cite{jentschura:11}), we use the
estimation of uncertainty from Ref.~\cite{mohr:16:codata}, which is about three times larger than
the one of Ref.~\cite{jentschura:11},
\begin{align}
\Delta \nu_{6} = -5.38\,(35)~\mbox{\rm kHz}\,.
\end{align}

The final theoretical value of
\begin{align}
\Delta f_{\rm th} = 670\,999\,567.88\,(42)~\mbox{\rm kHz}\,,
\end{align}
replaces the previous result $\Delta f_{\rm th}(\mbox{\cite{jentschura:11}}) =
670\,999\,566.90\,(89)~\mbox{\rm kHz}$. Combining the theoretical value $\Delta f_{\rm th}$ with
the experimental result from \cite{parthey:10,jentschura:11}, we obtain the updated result for the
mean-square charge-radii difference
\begin{align}
r^2_d - r^2_p = 3.820\,70\,(31)\,\mbox{\rm fm}^2\,,
\end{align}
which agrees with but is twice as accurate as the previous value of $3.820\,07\,(65)$~fm$^2$
\cite{jentschura:11}.

\end{document}